\documentclass[11pt]{bmc_article}    
\pdfoutput=1

\usepackage{cite} 
\usepackage{url}  
\usepackage{ifthen}  
\usepackage{multicol}   
\usepackage[latin1]{inputenc} 
\usepackage{graphicx}
\usepackage{amsmath}
\usepackage{multirow}

\newboolean{publ}

\newenvironment{bmcformat}{\baselineskip14pt\sloppy\setboolean{publ}{false}}{\baselineskip20pt\sloppy}

\begin{document}
\begin{bmcformat}


\title{Transcription Factor-DNA Binding Via Machine Learning Ensembles}
 

\author{Yue Fan$^1$%
        \email{Yue Fan - yue@bu.edu}
        and
        Mark Kon\correspondingauthor$^{1,2}$%
        \email{Mark Kon\correspondingauthor - mkon@bu.edu}
        and
        Charles DeLisi$^3$%
        \email{Charles DeLisi - @bu.edu}
        }


\address{%
    \iid(1)Department of Mathematics and Statistics, Boston University, 111 Cummington St., Boston, MA 02215\\
\vskip 2pt
    \iid(2)Department of Mathematics, MIT, 77 Massachusetts Ave., Cambridge, MA 02139\\
    \iid(3)Bioinformatics Program, Boston University, 24 Cummington St., Boston, MA 02215}%

\maketitle


\begin{abstract}
\subsection*{Motivation:}
The network of interactions between transcription factors (TFs) and their regulatory gene targets governs many of the behaviors and responses of cells. Construction of a transcriptional regulatory network involves three interrelated problems, defined for any regulator: finding (1) its target genes, (2) its binding motif and (3) its DNA binding sites. Many tools have been developed in the last decade to solve these problems. However, performance of algorithms for these has not been consistent for all transcription factors. Because machine learning algorithms have shown advantages in integrating information of different types, we investigate a machine-based approach to integrating predictions from an ensemble of commonly used motif exploration algorithms.
        
\subsection*{Results:}
We have developed an ensemble methods in a machine learning (ML) framework that combine predictions from five known motif and binding site exploration algorithms.  For a given TF, the ensemble starts with position weight matrices (PWM's) for the motif, collected from the component algorithms.  The collected ensemble of PWM's is used as a dimension reduction tool, identifying significant {\it PWM-based subspaces} for analysis. Within each subspace a machine classifier is built for identifying the TF's gene (promoter) targets (problem 1 above).   These  PWM-based subspaces form an ML-based sequence analysis tool, particularly useful in small sample situations.  Problem 2 (binding motifs) is solved by agglomerating k-mer (string) features PWM-based subspaces that stand out in identifying gene targets. We approach Problem 3 (binding sites) with a novel native machine learning approach, the  \textit{w}-scanning model.  This uses each gene promoter's string features and their ML importance scores in a classification algorithm to locate binding sites across the genome.

For target gene identification, averaged over 88 yeast TFs, this method improves performance (measured by the F1 score) by about 10 percentage points over the (a) motif scanning method and (b) the coexpression-based association method.

For binding motif identification, the top motif predictions from this method are reasonably similar to the known motifs for 62 out of 88 TFs, which outperformed 5 component algorithms as well as two other common algorithms (BEST and DEME).

For identifying individual binding sites on a benchmark cross species database (Tompa et al., 2005) of 56 TFs, this method achieved a similar performance to the best performer without much human intervention. It also improved the performance on mammalian TFs, for which the training sample size is often larger.

\subsection*{Conclusion:}        The ensemble can integrate orthogonal information from different weak learners (potentially using entirely different types of features) into a machine learner that can perform consistently better for more transcription.  The TF gene target identification component (problem 1 above) can be particularly useful in constructing of a more complete transcriptional regulatory network from a smaller sub-network based on known TF-target associations.  The ensemble is easily extendable to include more tools as well as future PWM-based information, from possibly new motif algorithms.

\end{abstract}

\ifthenelse{\boolean{publ}}{\begin{multicols}{2}}{}


\section*{Introduction}
The expression of genes in tissue and biological pathways is primarily controlled by transcriptional regulation \cite{gerstein07, Davidson, tijan_enhancer}. Transcription factors (TFs, also known as regulators) are proteins that initiate transcription of a gene (the target) by binding, with different levels of affinity, to nearby binding sites, usually 6-15 base pairs wide. These sites are often located in DNA within so-called promoter regions that are hundreds to thousands of base pairs long, usually upstream of the gene \cite{tijan_enhancer,Bulyk2008,Dai2007}. Current problems have included identification for every known regulator,  of (1) its target genes, (2) its binding motif (the general DNA pattern to which it binds) and (3) its DNA binding sites or locations. As genomes have been sequenced, new methods have provided significant results toward solving these problems\let\thefootnote\relax\footnotetext{This work was partially supported by NIH grant 1R21CA13582-01 and NIH grant 1R01GM080625-01A1}.

Current methods \cite{Consensus, annspec, AlignACE, BioProspector, MEME95, MEME, MDscan, Gibbs93} typically begin by identifying motifs for a given TF, searching for common DNA patterns in a collection of promoter regions of known or suspected target genes (we denote these as the \emph{positive set}). A binding motif is usually represented as a position weight matrix (PWM) \cite{Stormo1990,Stormo1999,Stormo2000} whose $j^{th}$ column consists of the four probabilities of the DNA bases $A$, $C$, $G$, and $T$ in position $j$ of the motif. The motif can then be used to detect new target genes and corresponding binding sites via rescanning of its PWM through the promoter regions of candidate genes \cite{Stormo1990,Stormo1999,Stormo2000}. The scores (\emph{scanning} or \emph{rescanning scores}) for each position in the promoter score it against the probability distribution defined by the PWM.  A promoter position with a rescanning score higher than a given threshold is reported as a new binding site and suggests a new gene target for the given TF.

More recently high dimensional machine learning methods have been
used for such analysis and have in some cases achieved state-of-the-art performance  
\cite{Dustin2006, leslie, SVMotif}. Several algorithms which form a suite for investigating transcription regulation have been developed based on a common machine learning framework \cite{Dustin2006, SVMotif}. In this paper, we introduce an ensemble machine method built on this framework, integrating information from diverse motif discovery algorithms.

Because of the nature of machine learning, the framework is initially based on solving a classification problem, in this case separating gene targets from non-targets (of a given TF) in a training or test data set. To determine whether a gene is a target, its promoter region is mapped into one or more high dimensional feature spaces, using maps capturing promoter properties that determine TF binding. A common and powerful feature space is the so-called  $k$-mer spectrum (or string) space.  This represents a gene $g$ (actually its promoter) in terms of a vector whose components are counts $k$-mers (short consecutive DNA strings of length $k$) in the promoter region \cite{Dustin2006, SVMotif}. For a fixed TF, the feature map $\phi(S_i)={\mathbf x_i}$  takes the promoter sequence $S_i$  of a potential target gene $g_i$ into a feature vector ${\mathbf x_i}\in F^{(k)}$ (the \textit{$k$-mer feature space}), whose $j^{th}$  component  $x_{ij}$ counts occurrences in $S_i$, of the $j^{th}$ $k$-mer (in an indexed list of all $k$-mers).
Each sequence $S_i$ is also labeled as $y_i=\pm 1$ indicating whether it represents a regulatory target (positive) or a non-target (negative). 
Target/non-target data are typically determined under a union of different experimental conditions.

A machine learning classifier function $f(\mathbf x)$ is trained on the dataset of known target/nontarget genes $D = \left\{\mathbf x_i, y_i\right\}_{i=1}^N$. The function $f(\mathbf x)$ is selected so that for training examples $\mathbf x_i$, the value $f(\mathbf x_i) >0$ for positive samples $\mathbf x_i$ (where $y_i =1$), and $f(\mathbf x_i) < 0$ for negative ones ($y_i = -1$).  The trained $f(\mathbf x)$ is then used to classify a new gene with promoter $S$ as positive (target) or negative (non-target), based on the classification $f(\mathbf x)$ of feature vector $\mathbf{x}=\phi\left(S\right)$; see \cite{Dustin2006}).  The ML algorithm to train $f(\mathbf x)$ here is the SVM - similar procedures have been used to classify proteins via amino acid sequences
\cite{leslie}.

As mentioned, an important goal of DNA binding analysis is to understand the transcriptional regulatory relationships between TFs and genes, with the purpose of identifying a new or extending a known regulatory gene network.  Note that in the machine learning framework, identifying TF regulatory target genes does not require a prior choice of motif.  To identify new targets of a TF, the classifier $f(\mathbf x)$  can be used directly \cite{Dustin2006}. 

Identification of new TF gene targets using training sets of known targets has previously been based on two common  methods: the above-mentioned motif scanning method \cite{Stormo2000} and coexpression-based association analysis \cite{CLR}.  The latter method and other machine learning methods for identifying targets have been shown to use information more efficiently and achieve better gene classification performance \cite{Dustin2006,vert2008}.

The strongest  $k$-mer  string features used by our gene target classifier (for a given TF) are usually related or identical to the  $k$-mer strings in DNA that in fact bind the TF.  Based on this fact a tool, SVMotif, was developed to search for binding motifs (PWMs) by agglomerating the most discriminative $k$-mers into binding motifs \cite{SVMotif} within a structured machine learning framework.  Specifically, for a fixed length $k$, the discriminative $k$-mer features (separating targets and non-targets) will overlap to form a longer (motif) sequence that can be used to form a candidate PWM.

Tested on the yeast genome, this method was able to discover 57 true motifs  out of 100 TFs among its top three predictions per TF \cite{SVMotif} - this included 49 ungapped motifs and 8 gapped motifs. The results were comparable to or better than two other commonly used algorithms, AlignACE and BioProspector. In contrast to many conventional motif-finding methods, SVMotif uses a set of known or hypothesized negatives (non-target genes) in addition to positives to provide more specific genomic background information. For choosing negative examples, randomly selected sequences can be used as representatives of the genome-wide background. Alternatively, simulated sequences can also be used to mimic a statistical background sequence model.

A support vector machine (SVM) classifier has the form of a linear discriminant, i.e., $f(\mathbf x) =  \mathbf w \cdot \mathbf x + b$, predicting that the gene with feature vector $\mathbf x$ is a positive (target) if $f(\mathbf x) > 0$.  The  $w$-vector $\mathbf w$ has components $w_i$ that are largest when the feature $x_i$ (counting occurrences of the $i^{th}$ listed $k$-mer in the promoter) is most significant in discriminating whether the gene is a target.  Thus the magnitude $|w_i|$ can also score the likelihood that the $i^{th}$ $k$-mer appears in the binding motif, helping to construct $M$. Since this identifies the most likely binding  $k$-mers, it can also be used to score potential binding sites of the TF in the promoter. This can be done by scanning the promoter and scoring each location using the $w$-vector directly, instead of the derived PWM. We will describe this $w$-scanning method and test it on a well-known benchmark dataset \cite{Tompa05}.

An analysis of 14 existing motif discovery algorithms \cite{Tompa05} suggested that no single algorithm can perform consistently well for every transcription factor. To take advantage of different strengths, ensemble methods, which combine predictions from different algorithms, have been of interest for refining predicted motifs. One type of ensemble approach scores motifs in an candidate pool by measuring their ``goodness". WebMOTIFS (the web interface of TAMO; \cite{WebMotif, TAMO}) assesses each candidate motif with several statistics, (e.g. hypergeometric enrichment score; \cite{Harbison04}), and a ranked list is reported. 

An existing ensemble approach for motif finding \cite{Hu06,Reddy1,BSG,Chen09} starts with multiple predicted binding sites of the TF based on several algorithms. The locations agreed to by the most algorithms are reported as binding sites.  The motif matrices are then formed by agglomerating the subsequences at those locations into a PWM. A scoring scheme is then used to rank predicted motif matrices based on their information content \cite{PAVESI04} and their matching frequency in the positive set. Some algorithms then optimize their motif score by locally adjusting binding locations to improve accuracy \cite{BEST,GAME,BioOptimizer}. One ensemble to which we will compare our algorithm, BEST \cite{BEST}, is of this type. At the end the algorithm reports motifs and the subsequences used to form them. Such ensemble algorithms can show a huge improvement over their individual components. Nevertheless, because these ensemble methods are used only to select only a single `best' choice of motif that is ultimately used, the out-of-training TF target and binding site identification can still have a high false positive rate if conventional PWM scanning methods are used \cite{Bindingsite} in the ensemble.

In this paper, we use the above-mentioned ML representations of
candidate promoter targets of a TF to develop a modular and
extendable ensemble machine framework, SVMotif Ensemble. Using this
we develop approaches for all three of the above problems
(identifying (1) target genes, (2) binding motifs and (3) binding
sites), within this framework, and improve performance for all three.
In particular the approach produces an ensemble-based classifier for out-of-training identification of TF target genes and binding sites, replacing commonly used PWM scanning models. Here we will focus more on the framework of this algorithm rather than selecting and tuning its individual components to obtain the most accurate predictions.  Thus only five widely-used algorithms, BioProspector \cite{BioProspector}, AlignACE \cite{AlignACE}, MEME \cite{MEME,MEME95}, Weeder \cite{Weeder2001} and SVMotif \cite{SVMotif} are integrated, without any fine-tuning of their combinations.  The resulting ensemble is fully scalable to allow other motif discovery tools to be added as future components.

We mention several points about the algorithm's properties as related to the above problems.

(1) In applying the algorithm we first validated the ensemble machine on discovery of \textit{S. cerevisiae} transcriptional  regulatory gene targets. For target gene identification, averaged over 88 TFs, this method improves precision and recall (measured by the $F_1$ score) by about 10 percentage points over the (a) motif scanning method using PWMs generated from 5 aforementioned individual algorithms. 

We also compared the ensemble as a gene target identifier with (b) the coexpression-based association method \cite{vert2008} and (c) a previously developed machine-based $k$-mer method \cite{Dustin2006, Dustin2008}. The ensemble produces improvements, varying over different transcription factors.

(2) We have used the same dataset as in (1) for testing performance of SVMotif Ensemble in binding motif identification; the top motif predictions are highly similar to the known motifs  for 20 out of 88 TFs and reasonably similar to the known motifs for another 42 out of 88. SVMotif Ensemble not only outperformed each of its 5 component algorithms, but also outperformed another ensemble algorithm - BEST \cite{BEST} and another discriminative method - DEME \cite{DEME}.

(3) We also tested the function of identifying individual binding sites on a standard benchmark database \cite{Tompa05}, containing 56 datasets from the human, mouse, \textit{D. melanogaster}, and \textit{S. cerevisiae} genomes. The ensemble is comparable to the best performer - Weeder with a special ad-hoc binding site selection producure \cite{Tompa05} as measured by both nucleotide and site-level performance measures (See \cite{Tompa05} for measure definitions). In addition, for mammalian datasets, which usually contain more training examples of TF-gene interactions (such large datasets are becoming much more prevalent in current research; see \cite{ENCODE}), there is an improvement as well over Weeder in identification of binding sites.

Some methodological points here are worth noting. The first involves a connection between machine learning classifiers like $f(\mathbf x)$ (here the SVM linear discriminant) and PWM rescanning-based classifiers, for both TF target gene identification and binding site identification (see Section \ref{sec:method.2}) .  This connection essentially will show SVM-based TF target gene classifiers to form a superset of the set of gene classifiers which are based on PWM re-scanning (see Section \ref{sec:method.2}).  Based on this, it will be possible to conclude that such machine classifiers form a strictly better alternative to PWM rescanning in solving out-of-training identification problems (for both TF target gene and binding site identification).  Specifically, consider the SVM-based classifier. Let $f(\mathbf x)=\mathbf w \cdot \mathbf x + b$ be the SVM-based gene classification function (for a given TF), for $\mathbf x$ in the  $k$-mer feature space $F$. On one hand, note that the coefficient $\mathbf w=\nabla f$ contains the information used as related the TF target gene problem, and hence the motif finding problem.  We will show (Section \ref{sec:method.2}) that, given any PWM $M$ for the TF, it is possible to emulate $M$-based TF target gene classifications using a classifier  $f(\mathbf x)=\mathbf w \cdot \mathbf x + b$ whose classifications of genes $g$ are effectively identical to those based on motif $M$.  To this extent we can consider the machine classifier  $f$ to be a strict generalization of the PWM scanning classifier.  

In particular the set of vectors $\mathbf w$ giving classifiers $f$ that are compatible as above with rescanning by a fixed motif $M$ form only a subset of {\it possible} $w$-vectors.  The latter can be searched in constructing all SVM candidates $f(\mathbf x)=\mathbf w \cdot \mathbf x + b$, from which the best SVM is selected.  Given that the SVM optimizes a loss function (based on errors in the training samples) by searching a collection of classifiers larger than just those based on PWM, it should be able to perform at least as well as any PWM-based algorithm. We believe that if sample size is large enough the SVM algorithm generalizes significantly better than PWM methods to predict out-of-training targets of a TF.

The second point involves an advantage of the ML approach to ensemble learning, based on the nature of feature space $F$.  This space (here consisting of $k$-mer count feature vectors) provides a common source-independent framework into which information from component algorithms is imported.  We note also that its dimension can be reduced dramatically based on the ensemble technique.  Each candidate PWM $M_j$ obtained from component PWM algorithms ensemble can be used to generate a reduced low-dimensional subspace $F_j$ of $F$, the \emph{PWM subspace}. This is obtained by extracting from the PWM a selection of $k$-mer features \cite{SVMotif}  which are most likely based on the PWM.  The reduced space $F_j$ is the span of these $k$-mer features. This dimension reduction approach in our ML ensemble framework is initially used for identifying TF target genes.  The choice of PWM subspace $F_j \subset F$ carries the information in $M_j$ to the feature space $F$.  Training can be done on a reduced machine classifier $f_j(\mathbf x)$ on subspace $F_j$ (yielding a coefficient vector $\mathbf w$ within $F_j$), derived from the ensemble via $M_j$.  Thus each ensemble motif algorithm is a component dimension reduction tool producing subspaces $F_j\subset F$.  We will call the map from PWM $M_j$ to subspace $F_j$ a \emph{subspace-valued weak learner}, weak because the resulting dimension reduced mapping $f_j$ is a relatively small component of the full learning algorithm producing the machine classifier $f(\mathbf x)$ in $F$. We remark that the discrimination power of subspace $F_j$ in finding TF gene targets is a good measure of the quality of the PWM $M_j$ itself.--  it provides an alternative to the area under ROC curve (AUROC) score of a PWM measuring binding site identification, first introduced by \cite{Clarke}.

Third, finding gene targets using the (SVM-derived) linear discriminant $f(\mathbf x)$ extends directly to a method for finding specific binding sites, extending the standard PWM scanning binding site identification method.  We call this method \emph{$w$-scanning}; it uses the above feature space $F$ to find binding site positions. Like standard PWM scanning, the approach scans a promoter by scoring each of its $k$-mers using the above SVM $w$-vector.  As mentioned above, this strictly extends capabilities of PWM-type scanning methods, in particular avoiding the implicit assumption that binding site positions are independent.  This has a two-sided effect. On one hand, if the independence assumption is invalid, $w$-scanning can improve accuracy over PWM-based models. However, the approach needs a relatively large training set of known positives, because learning complexity is higher. In particular the method may overfit noise (false dependences among motif positions) when trained on small samples sizes.

Last (but not least), the machine learning approach easily combines information sources that go beyond sequence information. This can include information like experimental mRNA co-expression, phylogenetic sequence conservation, and nucleosome positioning. In \cite{Dustin2006}, such information was combined with $k$-mer features to find TF-target associations.  Thus the ensemble can be expanded to include PWM information from new algorithms, as well as other sources such as gene expression and sequence conservation.

Our approach uses each component PWM algorithm to provide candidate motifs/PWMs.  Each PWM $M_j$ generates a large number of strings `typical' for it, that then form the basis for an associated {\it synopsis subspace} $F_j$ of the full string space $F$.  Machine training the set of positive and negative promoters just on $F_j$ yields for each test promoter's feature vector $x$ a so-called \textit{synopsis score} $f_j(\mathbf x) = \mathbf w^{(j)}\cdot \mathbf x^{(j)} + b_j$.  This is the SVM discriminant score based only on $F_j$.  The ensemble of individual scores $f_j(\mathbf x)$ themselves form a reduced feature vector with one feature (the value $f_j(\mathbf x)$) for each $M_j$, giving an extensive dimension reduction from $F$.  This reduced vector with $j^{th}$ component $f_j(\mathbf x)$ is called the {\it synopsis vector}, and the space of these is denoted as the {\it synopsis space}.  We should mention that other ways to combine machine learning information can be used instead - these include adding kernels corresponding to different subspaces $F_j$ (kernel addition), and forming direct sums of the feature spaces $F_j$ \cite{EnsembleConf}. Compared with these, however, this method is computationally efficient.  Combined with a sub-feature selection tool, (selecting only important synopsis features $f_j(\mathbf x)$), this maintains scalability, leaving room for more useful future information.

SVMotif Ensemble is a machine learning software suite for solving the above problems. It takes known target and non-target promoter sequences (the training set) as input, and automatically runs the input algorithms of the ensemble (e.g. Bioprospector [ref], Alignace [ref] to get potential PWMs. These PWM's are then used to reduce dimensionality of the full machine learning string feature space $F$.  On the reduced spaces $F_j$, classifier functions $f_j({\mathbf x})$ are trained to distinguish target and non-target promoters, forming the trained ensemble machine.  

As a suite for transcription regulation analysis, the trained SVMotif Ensemble predicts target genes of the TF, outputs a binding consensus motif matrix, and predicts potential binding sites near each target. The user can store the learned ensemble machine, which contains the learned subspace information as well as information from training samples, for future use. Compared to the traditional way of using PWMs as the direct information source, the ensemble contains more and more accurate information for our identification problems. The software suite is available for download from our website at \url{http://cagt10.bu.edu/SVMotif}.

\section*{Results}
\label{sec:result}
\subsection*{Experimental Protocol}
\label{sec:result.1}
To identify gene targets of a given TF, we used a benchmark dataset of TF-DNA interactions from \cite{Dustin2006} that contains positives (known gene targets) and negatives (genes with large $p$-values in ChIP-chip experiments; \cite{Lee_Network02}), based on information for 163 yeast transcription factors. We also downloaded PWMs for 102 TFs (out of the 163) that are available from the UCSC database \cite{UCSC10}, to test against performance of the PWM scanning model, using known PWMs. We excluded those TFs with less than 20 known targets in our dataset since ML performance is unreliable for small numbers of positive examples.  This left 88 TFs to be tested. For each TF, we selected all known positives (targets) and an equal number of (presumed) negatives as our experimental dataset. A 5-fold cross validation was performed on each dataset (for an individual TF), dividing the target/non-target genes into 5 equal groups.  Specifically, to ensure full isolation of training and test data (failure to do this would overstate performance measures), the promoter sequences (including positives and negatives) were randomly divided into 5 portions. For each withheld test data fold, we used the remaining 4 folds of data to train individual weak learners (here using AlignACE, BioProspector, MEME, SVMotif and Weeder without the special ad-hoc procedure) and 3 different ensemble methods, and tested the resulting target gene classifiers on the withheld test fold. This was repeated withholding all 5 folds (as test folds) one at a time, and we obtained cross-validation predictions for all genes in the dataset.  The SVM used output probability-values between 0 and 1 (probability of membership in one of the classes), which were used in the scores \cite{Platt}. The $F_1$ score, defined by
\[ F_1 = 2\cdot\frac{Recall \cdot Precision}{\left(Recall + Precision\right)} \] was used as an overall measure of prediction quality. Performance is discussed in section \ref{sec:result.2}

For the second task of identifying binding motifs, the same 88 yeast transcription factors were tested.  We used all known positives and a randomly selected equal number of negatives to train both weak learners and ensembles. As a performance measure, we calculated the motif similarity (Section \ref{sec:method.6}) between the UCSC PWMs and each of the top 3 predictions from all of the tested ensemble algorithms. Performance is discussed in Section \ref{sec:result.3}.

For the final task of identifying binding sites, we chose benchmark TF binding site datasets from \cite{Tompa05}. These datasets covered 4 species, including human, mouse, \textit{Drosophila melanogaster} and \textit{Saccharomyces cerevisiae} data. Only positive sequences were originally provided in each of the datasets.  For a training set of negative (non-target) genes, we downloaded 1000 base pair upstream sequences. This data was obtained for each of the four species from the whole-genome database from the UCSC genome browser. Randomly selected sequences from this collection were used as negatives. Because the number of positive sequences was small, we selected twice as many negatives as positives for training the ensemble classifiers. Performance is discussed in Section \ref{sec:result.4}.

As mentioned earlier, five commonly used motif exploration algorithms were combined as weak learners - these were AlignACE \cite{AlignACE}, BioProspector \cite{BioProspector}, MEME \cite{MEME}, Weeder \cite{Weeder2001} without any ad-hoc binding site selection procedure as well as the SVMotif algorithm \cite{SVMotif} based on the full $k$-mer feature space. We selected top-ranked PWMs from each algorithm based on their own ranking scores. The selected candidate pool of motif matrices contained the top 5 motifs from AlignACE; the top 2 motifs from each different run of BioProspector (each based on a single motif width ranging from 7 to 12, with 12 PWMs in total); the top motif from separate runs of MEME, each using a different width ranging from 7 to 12 (6 PWMs in total); the top 5 motifs from SVMotif, and the top motif from Weeder. Since Weeder can only output individual DNA strings rather than PWMs, an in-house string agglomeration algorithm was applied to build PWMs. This setup yielded 29 motif matrices as the candidate pool for each transcription factor.

\subsection*{Ensemble Classification of Gene Targets}
\label{sec:result.2}
To benchmark prediction quality of gene targets of a TF, we first tested the performance of individual learners in the ensemble. Because the component algorithms generally output binding motifs\footnote{They also predict binding sites for training samples. However, this cannot be used to identify target genes in new samples.} rather than gene targets, we combined motif predictions with the PWM scanning model to identify TF binding targets (see Methods), based on aggregated PWM predictions  of the weak learners.  For component algorithms with multiple PWM predictions, we selected the PWM predictions whose scanning scores did best at identifying binding sites on the training portion of the dataset.  As also observed in \cite{Tompa05}, no single learning method could perform consistently well for all transcription factors (Fig.~\ref{fig:Components}).

\begin{figure}
\centering
\includegraphics[width=\textwidth]{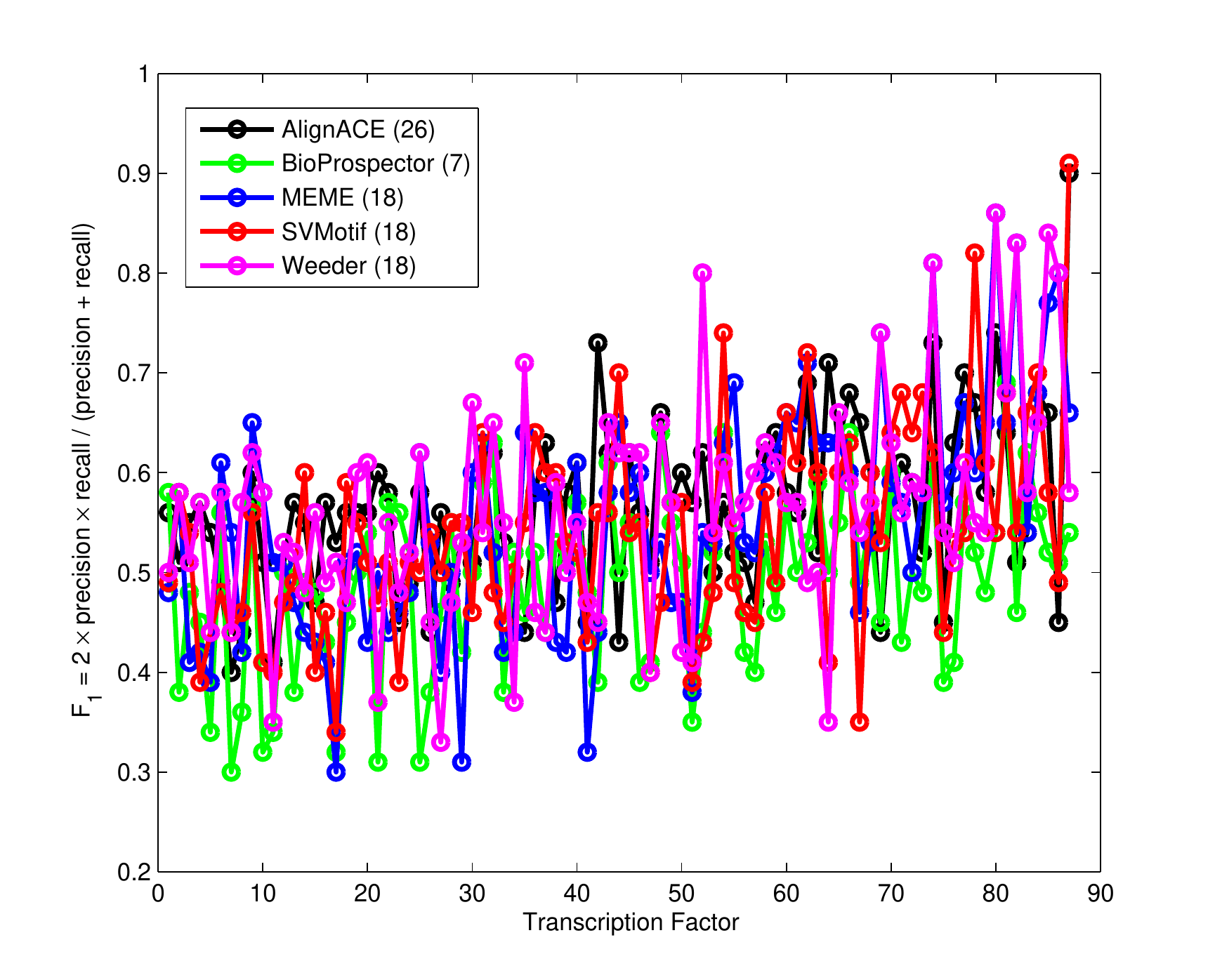}
\caption{Performance of component algorithms: Plot of $F_1$ score for each transcription factor and each component 
algorithm. The number in parenthesis after each component is the number of transcription factors for which the component showed the best performance.} 
\label{fig:Components}
\end{figure}

For comparison, we also tested another class of TF target identification algorithms, based on gene co-expression studies.  Such studies have been used as tools for gene regulatory network construction using various algorithms, for example in \cite{CLR} and \cite{vert2008}.  For such coexpression studies we use SVM algorithms in \cite{vert2008} (using coexpression databases in \cite{Dustin2006}) As pointed out in \cite{vert2008}; this method had previously performed reasonably well in predicting \textit{E. Coli} regulatory relationships. In this test, these expression-based classifiers achieved on average 57\% $F_1$ score on our yeast data.  

We also compared our ensemble algorithm with the previously developed SVMotif algorithm based on the full $k$-mer feature space, with $k = 6$. On the same dataset as above, the full $6$-mer space method achieved a 66\% $F_1$ score. In \cite{Dustin2006}, SVM performance using this full $k$-mer space $F$ dominated performances using other classes of feature spaces based on 25 other information sources, including co-expression data. However, the full $k$-mer space method can reach  computational limitations as the size of the motif becomes larger.

As mentioned the ensemble methods were tested on each TF. The average $F_1$ score (over 88 TFs) of the methods is approximately 10 percentage points higher than that of the best component algorithms (70\% versus 57\% for AlignACE; Fig.~\ref{fig:Ensemble}). 

\begin{figure}
\centering
\includegraphics[width=\textwidth]{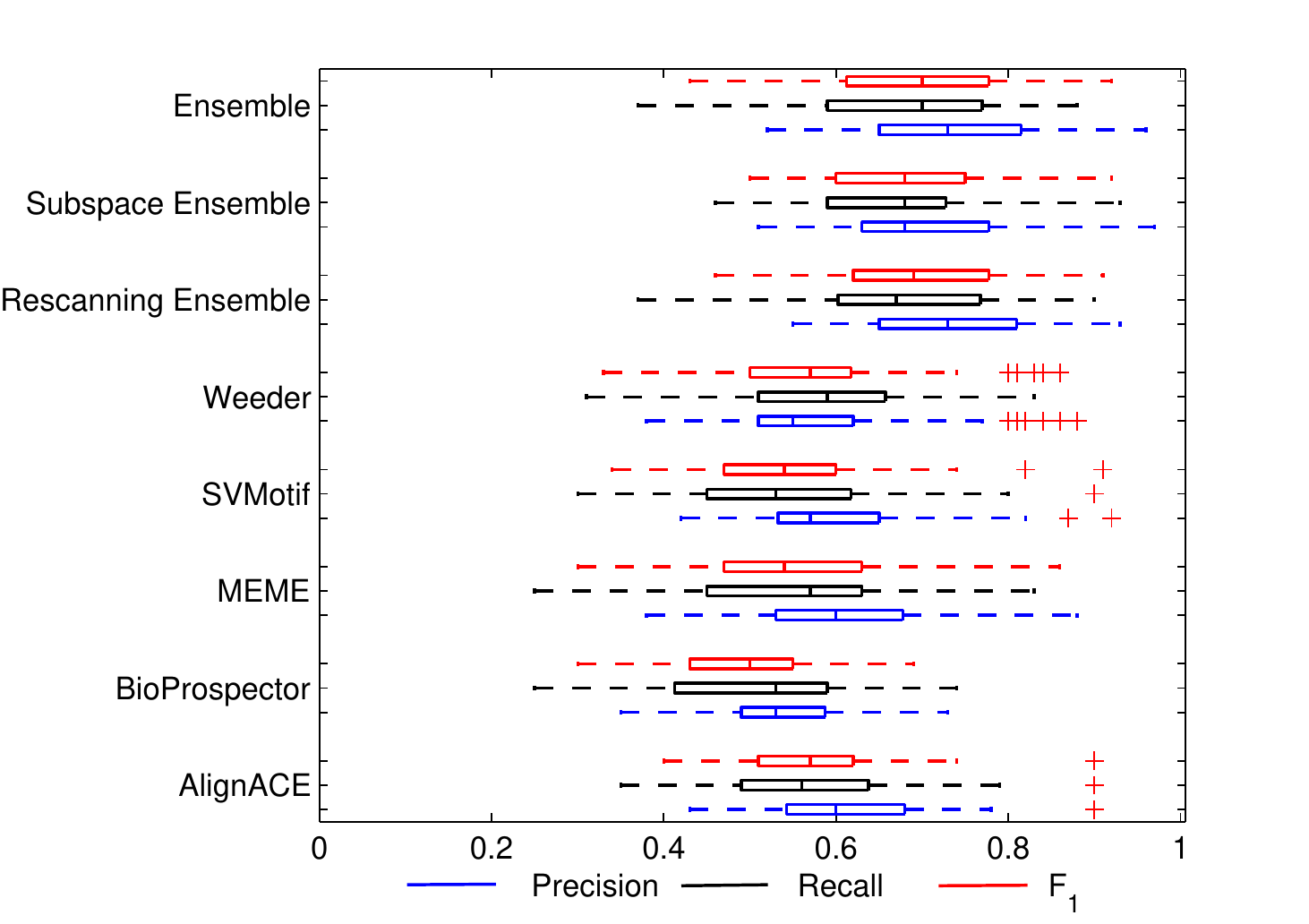}
\caption{Performance measures of ensemble algorithm versus individual components in gene target identification: 
	  Precision, recall and $F_1$ scores are computed for each transcription factor individually. This box-plot
	  chart shows the ensemble methods outperform other individual methods in $F_1$ score on the average by 10 percentage points. } 
	  \label{fig:Ensemble}
\end{figure}

Looking at performances on individual TFs, for 75/88 TFs the ensemble outperformed the best performing of its five component algorithms in TF gene target identification (Fig.~\ref{fig:Compare1}). Thus the integration of orthogonal (very different) algorithms can not only preserve best performance, but also improve overall performance.  
\begin{figure}[h] \centering 
\includegraphics[width=\textwidth]{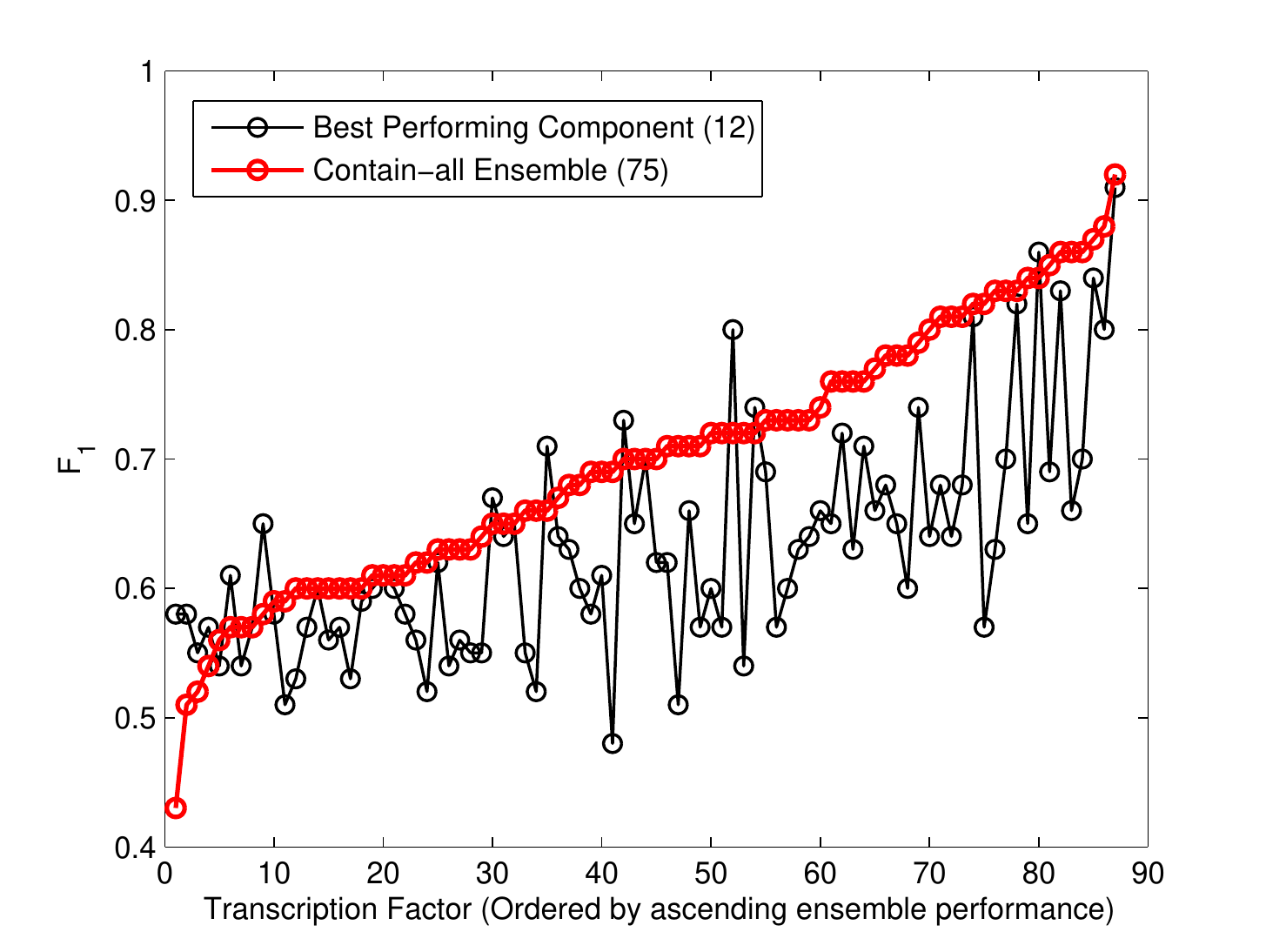} 
\caption{Ensemble versus best performing component: In terms of $F_1$ score, the ensemble method outperforms the best-performing individual component for 75transcription factors among 88 tested.} \label{fig:Compare1} \end{figure}

\subsection*{Ensemble Binding Motif Identification}
\label{sec:result.3}
Under a PWM rescanning model, any new identifications (of target genes or binding sites) rely on a good PWM estimate.  Though the machine classifier initially scores genes as potential TF targets and does not estimate a PWM, it is possible to build PWMs from the classifier by ranking and merging its most informative features affecting the SVM score. To assess the performance in predicting binding motifs, we computed the motif similarity (Section \ref{sec:method.6}) between our prediction and the UCSC \cite{UCSC10} standard motif $M_0$, for each TF.

We first considered similarities among PWMs in the candidate pool $\mathcal M$ (from the component weak learners) to the standard motif $M_0$. The best performing motif matrix among these (denoted as $M_b$) and its similarity scores to $M_0$ were used as an ensemble testing benchmark.  An ideal ensemble should reproduce this best performing matrix $M_b$ at the top of its list. We considered performance of both the top single and the top three ensemble predictions.  The top prediction and the top three predictions had similarity scores to $M_0$ that exceeded best individual component scores (see Section \ref{sec:method.6}) for 46 and 55 transcription factors, respectively (among 87 TFs tested, Fig.~\ref{fig:Motif1}). The top ensemble prediction also outperformed the top predictions from the another ensemble method, the BEST algorithm \cite{BEST}, and from another supervised method, DEME \cite{DEME}, for most transcription factors (Fig.~\ref{fig:Motif2}).

\begin{figure}[ht] \centering
\includegraphics[width=\textwidth]{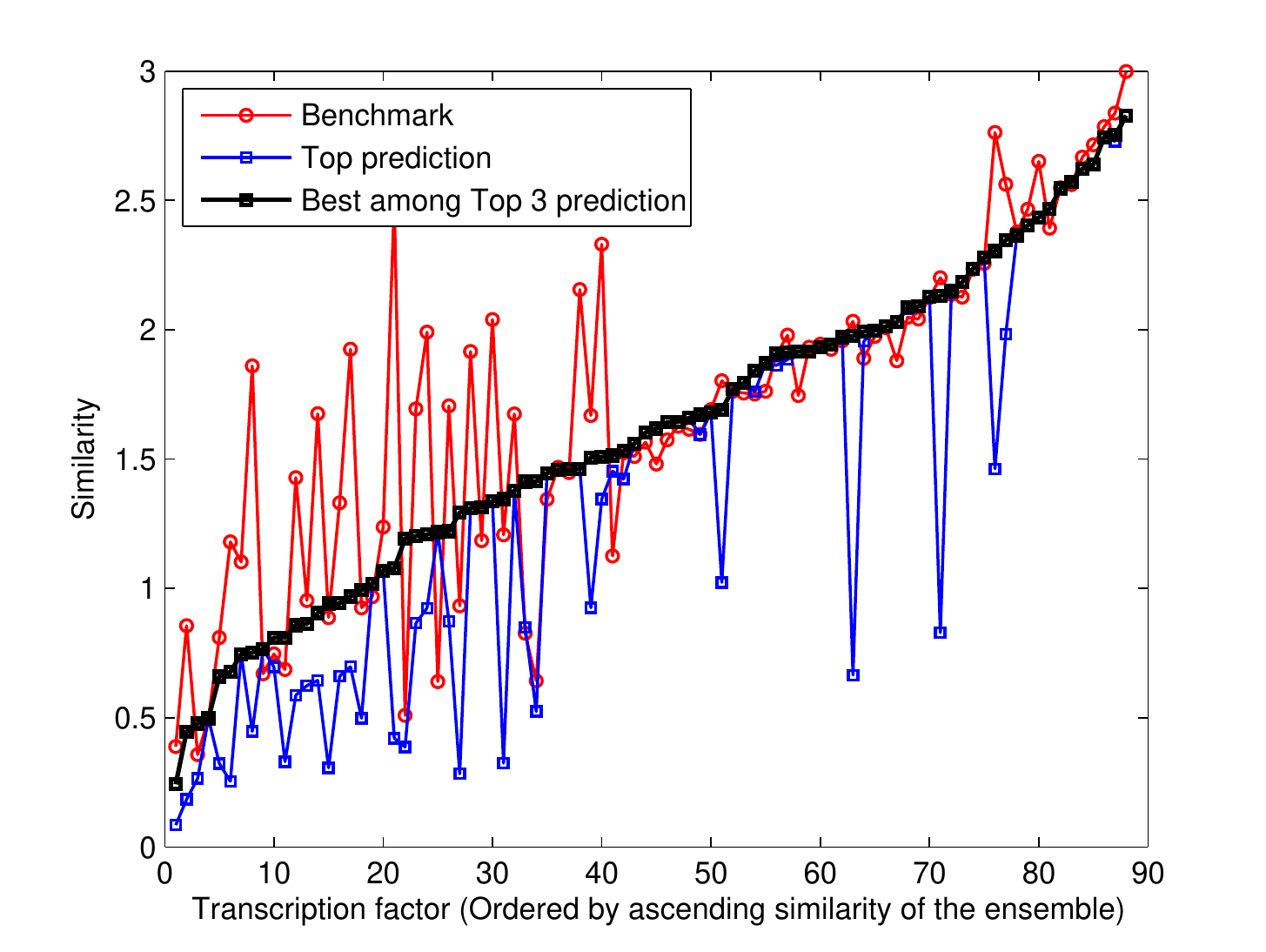} 
\caption{Selected motif similarities: ensemble vs. component weak learners. Benchmark denotes the PWM from any component algorithm with greatest similarity to the standard PWM $M_0$. Top prediction and best among 3 predictions refer to predictions of the ensemble algorithm. The top ensemble prediction was same as the benchmark for 46 out of 87 TF's.} 
\label{fig:Motif1}
\end{figure}

\begin{figure}[ht] \centering
\includegraphics[width=\textwidth]{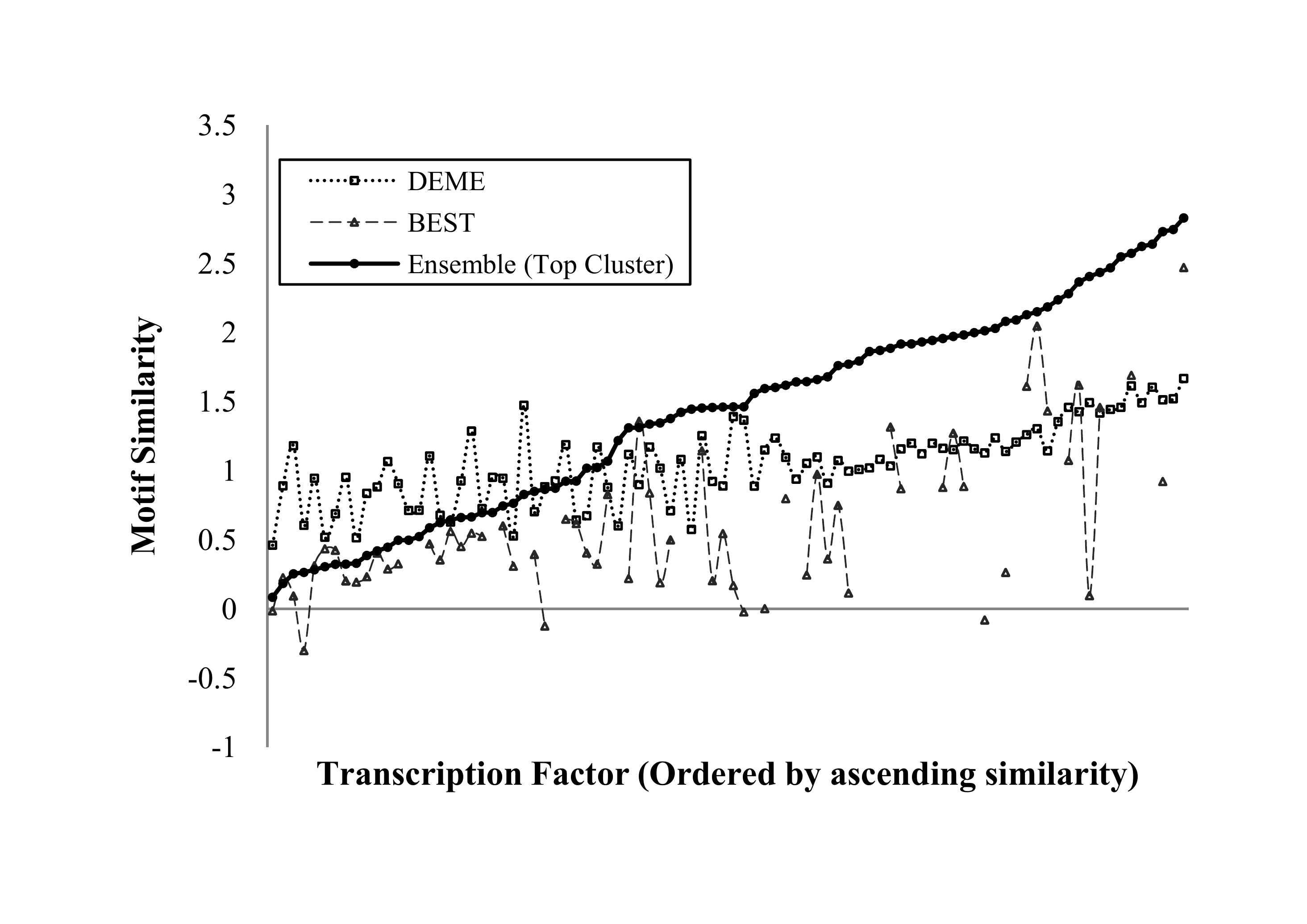} 
\caption{Selected motif similarities compared to BEST and DEME. Missing values for BEST indicate it did not output a result.} 
\label{fig:Motif2}
\end{figure}

To confirm our predictions were biologically meaningful we also looked at curated information from the Transfac database \cite{Transfac}, which reported binding motifs for 26 TFs out of the full list of 87. Among these, three of the weak learners (Bioprospector, AlignACE, and SVMotif) could jointly predict the correct motif for 21 TFs with their selected predictions (i.e., at least one of the three had the correct prediction among its pooled PWMs in 21 cases out of the 26). The ensemble method predicted the correct motifs for 18 out of these 21 cases as its top ranked prediction. This shows the ability of the subspace ensemble to integrate a variety of information from different weak learners and to pick out the most meaningful parts. If a sub-learner predicts the true motif, an ideal ensemble machine should give the right prediction, which occured in over 75 percent of these cases.

\subsection*{Binding Site Identification}
\label{sec:result.4}
We used the above datasets also to test how the ensemble method performs using sparse training sets (with small numbers of known gene targets). Since the number of training samples in such a benchmark is limited, a long motif pattern is difficult to detect. For example, if the true pattern is a  $10$-mer, this requires a learning machine to identify it within a $4^{10} \approx 1$ million-dimensional space of possibilities. Since the number of training data points (positives) might be as small as 10, identifying such a subtle signal in a high dimensional space is difficult. The present ensemble method collects PWM outputs from sub-learners using largely to reduce the search dimensionality to several dozen, with the effect of increasing the signal/noise ratio.

\begin{table}
\centering
\renewcommand{\arraystretch}{1.3}
\scriptsize
\
\begin{tabular}{cl}

\multicolumn{2}{l}{\textbf{Nucleotide Level}} \\ \hline\hline
\scriptsize{Notation}&\multicolumn{1}{c}{Definition and Formula} \\ \hline
$nTP$&number of nucleotide positions in both known sites and predicted sites \\ \hline
$nFP$&number of nucleotide positions not in known sites but in predicted sites \\ \hline
$nTN$&number of nucleotide positions in neither known sites nor predicted sites \\ \hline
$nFN$&number of nucleotide positions in known sites but not in predicted sites \\ \hline
$nSn$&Sensitivity / Recall: $nSn=\frac{nTP}{nTP+nFN}$                            \\ \hline
$nPPV$&Positive Predictive Value/Precision: $nPPV=\frac{nTP}{nTP+nFP}$         \\ \hline
$nSp$&Specificity: $nSp = \frac{nTN}{nTN+nFP}$                                   \\ \hline
$nPC$&Performance Coefficient: $nPC = \frac{nTP}{nTP + nFN + nFP}$                \\ \hline
\multirow{2}{*}{$nCC$}&Correlation Coefficient \\ 
 &\tiny{$nCC = \frac{nTP nTN - nFN nFP}{\sqrt{(nTP+nFN)(nTN+nFP)(nTP+nFP)(nTN+nFN)}} $} \\ \hline
$nF_\beta$&$F_\beta$ score: $nF_\beta = \frac{(1+\beta)\cdot nSn \cdot nPPV}{nSn+nPPV}$ \\ \hline \hline
\\
\multicolumn{2}{l}{\textbf{Site Level} \scriptsize{(``Overlap" indicates overlapping by at least 1/4 of a given site)}} \\ \hline\hline
\scriptsize{Notation}&\multicolumn{1}{c}{Definition and Formula} \\ \hline
$sTP$&number of known sites overlapped by predicted sites \\ \hline
$sFN$&number of known sites not overlapped by predicted sites \\ \hline
$sFP$&number of predicted sites not overlapped by known sites \\ \hline
$sSn$&Sensitivity / Recall: $sSn=\frac{sTP}{sTP+sFN}$ \\ \hline
$sPPV$&Positive Predictive Value/Precision: $sPPV=\frac{sTP}{sTP+sFP}$ \\ \hline
$sASP$&Average Site Performance: $sASP = \frac{sSn + sPPV}{2} $ \\ \hline
$sF_\beta$&$F_\beta$ statistics: $sF_\beta = \frac{(1+\beta)\cdot sSn \cdot sPPV}{sSn+sPPV}$ \\ \hline
\hline
\end{tabular}
\hspace{\baselineskip}
\caption{Definitions of performance metrics used in the assessment of motif discovery algorithms} 
\label{tab:ensemble.TompaMatrics}
\end{table}

The overall result of this test over the four species shows that the sensitivity of the ensemble method surpasses the best among those tested in \cite{Tompa05}. The best nucleotide-level sensitivity\footnote{Table \ref{tab:ensemble.TompaMatrics} lists the definitions of performance metrics used in \cite{Tompa05} and this paper} is below $10\%$ for all other algorithms tested, while the ensemble method gives $14.6\%$. Looking at site level sensitivity, our method has a value of $19.4\%$, indicating the ensemble successfully predicts about $20\%$ of true binding sites among the 4 species. The precision (a.k.a. PPV in \cite{Tompa05}) is still at a similar level to other algorithms. The $F_1$ score, which combines specificity and sensitivity, is comparable to the best component score, that of the Weeder algorithm with a special ad-hoc binding site selection procedure \cite{Weeder2001} (Fig.~\ref{fig:TompaA}). 

Because the ensemble starts with multiple motif models, it predicts more sites (signals along the promoter) in its first stage.  It then excludes false positives by restricting to just the primary motif cluster in the \textit{ad-hoc} analysis (Section \ref{sec:method.7}). From a dimension reduction point of view, the predictions in the first stage can effectively reduce the search space from the entire promoter region down to about 5 potential sites per sequence. The sensitivity of the ensemble is high, so the true site is more likely to be included in this initial list. In addition to sensitivity we also looked at the discovery power, measured by the proportion of TFs for which the algorithm has non-zero $F_1$ score. A non-zero $F_1$ suggests that the algorithm is able to predict at least one true binding site within the given datasets. Table \ref{tab:ensemble.TompaAll} shows the ensemble is able to identify at least one true binding site (non-zero $F_1$ score) for 22 out of 56 TFs. It performed the best among all algorithms tested. 

In order to distinguish the functioning from non-functioning binding sites, some \textit{ad-hoc} analysis is needed, such as a database search or conservation analysis. Experimental or computational approaches will also make sense under such circumstances. In addition, a more refined background model can also be used at this point (e.g., see discussion of Weeder's special ad-hoc procedure on the \cite{Tompa05} website) to select the most overrepresented binding sites. Such \textit{ad-hoc} analyses  will produce more useful results if the sensitivity (discovery power) of the computational algorithm is high.

In addition to our overall performance comparison using the 56 datasets based on cross-averaged statistics, we also compared the algorithms on these datasets individually. Using $F_1$ score as a metric we ranked 16 algorithms, including 14 tested in \cite{Tompa05}, alongside SVMotif and Ensemble, based on performance on each of the 56 TFs in the dataset. For each of the algorithms we counted the number of TFs for which its performance on binding site identification was ranked at the top, within the top 2 or within the top 3 algorithms. Among the 56 TFs, SVMotif Ensemble had the best performance on 11 and 9 of them, at nucleotide and site level respectively. It outperformed all other algorithms substantially (see Table \ref{tab:ensemble.TompaAll}).

We also noted that the ensemble method in particular performed better than other methods on mammalian datasets. As seen from Fig.~\ref{fig:TompaB}, both nucleotide level and site level $F_1$ scores surpass those of all other algorithms. Because of the nature of the machine learning approach, it performs better when sample size is relatively large. The correlation coefficient between the ensemble method's site level $F_1$ score and the number of positive examples per TF is $27\%$ for all four species, while the value for Weeder with special ad-hoc procedure is only $17\%$. Hence for a dataset with large sample size (say $> 100$), the machine learning method is more predictive. This is important given the large numbers of positive instances of TF binding sites obtained, e.g., in ENCODE  \cite{ENCODE}. 

\begin{figure}[ht] \centering
\includegraphics[width=\textwidth]{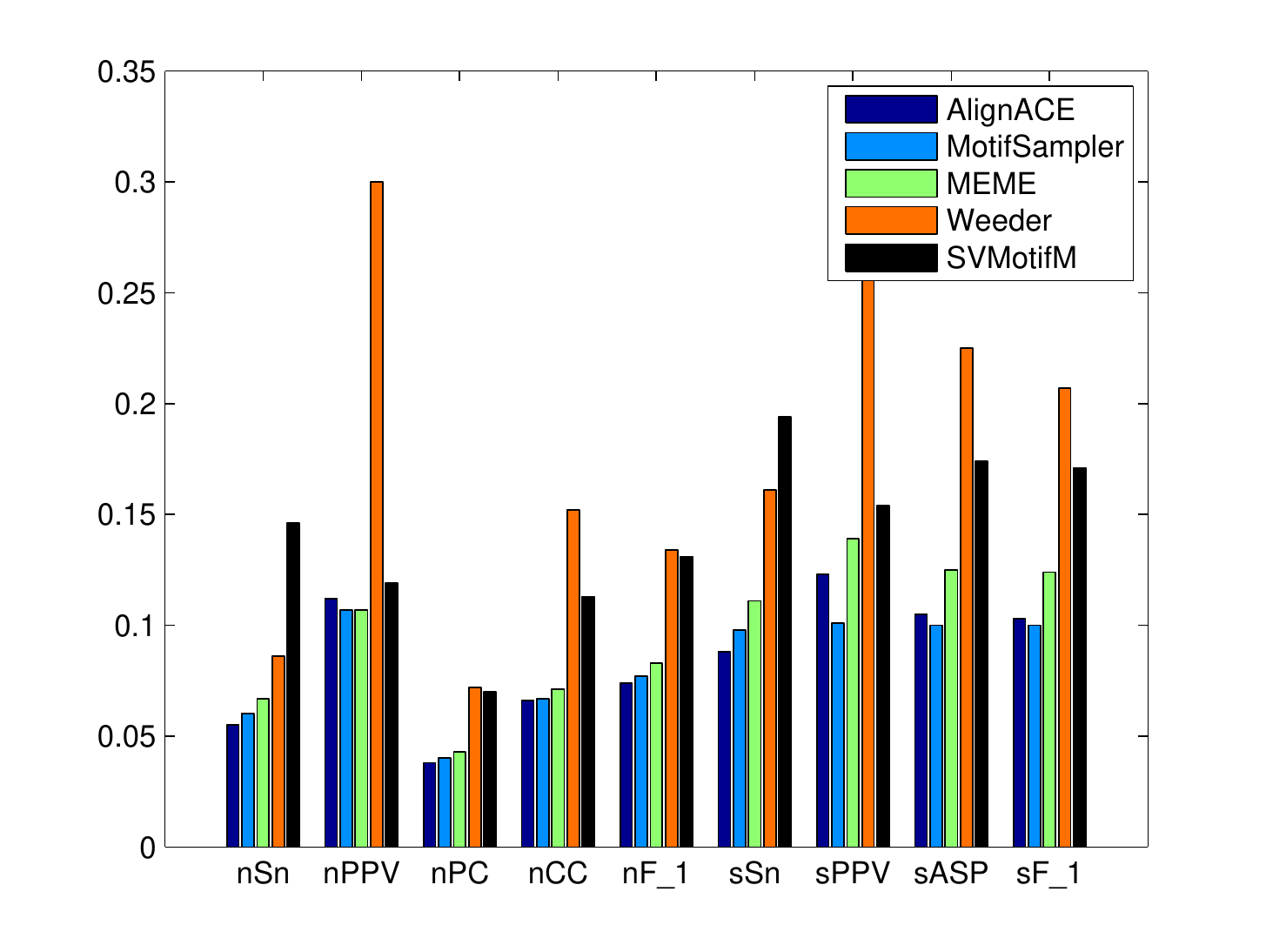} 
\caption{Performance metrics on all Tompa datasets for binding site identification.} 
\label{fig:TompaA}
\end{figure}

\begin{figure}[ht] \centering
\includegraphics[width=\textwidth]{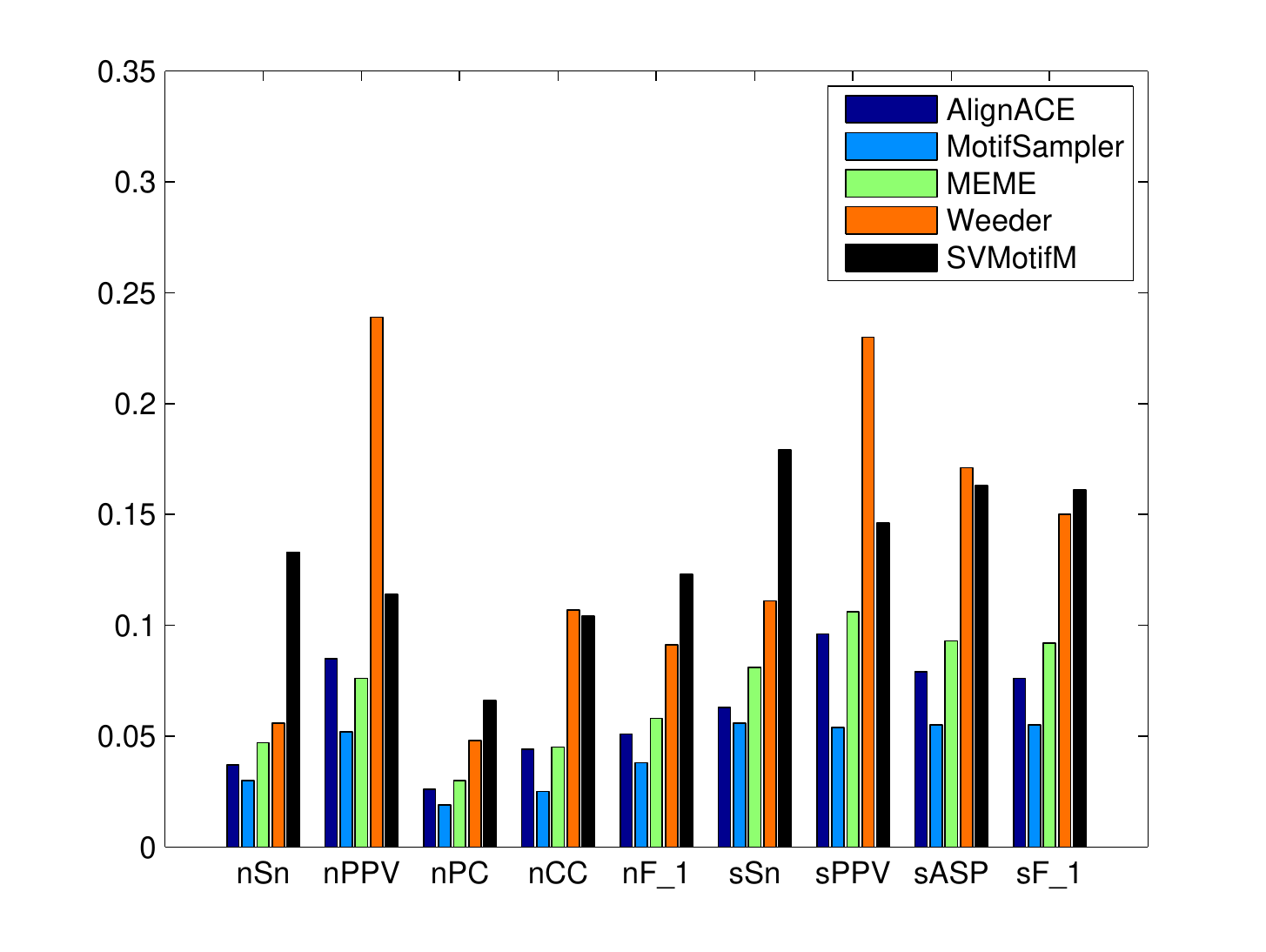} 
\caption{Performance metrics on all Tompa mammalian datasets for binding site identification.} 
\label{fig:TompaB}
\end{figure}

\begin{table}
\centering
\footnotesize
\begin{tabular}{l|cccc|cccc}
\hline\hline
\multirow{2}{*}{Algorithms}&\multicolumn{4}{c}{Nucleotide Level}&\multicolumn{4}{c}{Site Level} \\ \cline{2-9}
&\tiny{$F_1>$}0&Top&Top2&Top3&\tiny{$F_1>$}0&Top&Top2&Top3 \\ \hline
AlignACE&10&2&2&7&8&2&2&6 \\
ANN-Spec&25&5&8&12&20&2&6&7 \\
Consensus&7&1&1&2&6&1&2&4 \\
GLAM&15&0&2&4&11&0&3&5 \\
Improbizer&22&1&4&4&21&1&3&5 \\ 
MEME&23&3&7&12&20&5&9&13 \\
MEME3&19&3&5&7&15&3&5&8 \\
MITRA&16&1&3&4&10&1&4&5 \\
MotifSampler&21&4&8&10&18&4&7&11 \\
oligodyad-analysis&15&2&3&5&12&2&3&4 \\
QuickScore&14&1&5&6&7&0&1&1 \\
SesiMCMC&19&4&6&11&19&4&8&12 \\
WEEDER&18&4&6&7&18&7&9&11 \\
YMF&20&3&8&10&18&3&5&7 \\ 
SVMotif&20&5&11&17&19&2&6&8 \\ \hline
SVMotif Ensemble&24&11&15&16&22&9&12&13 \\ \hline
\hline
\end{tabular}
\hspace{\baselineskip}
\caption{Performance ($F_1$ scores) of 16 algorithms on individual Tompa datasets. Columns 2 and 6 indicate the number of datasets on which the algorithm has discovery power (i.e., produces non-zero nucleotide and site level $F_1$). Columns 3 to 5 and 7 to 9 indicate the number of datasets on which the algorithm performs best among the 16 (in terms of $F_1$ score).} 

\label{tab:ensemble.TompaAll}
\end{table}

\subsection*{The machine classifier is an alternative to the PWM model}
Identifying target genes of a TF is central to reconstructing regulatory networks, which has been of recent interest \cite{CLR, Lee_Network02, Harbison04, MacIsaac2006, Dustin2006}.  Up to now the majority of TF binding analysis has been based on PWM models. In conventional PWM models the rescanning score can be a powerful tool for identifying new target genes. As is shown in Section \ref{sec:method.2}, however, the PWM rescanning algorithm effectively forms a linear classifier, even when maximum local PWM scores are used.   A PWM cannot capture all information in its training sequences however, and the accuracy of such a classifier is not optimal on out-of-training samples. The machine classifier uses known target information directly and represents its decision rule in a high dimensional machine that effectively captures dependence information between bases. To compare the two models on gene target identification, we tested the PWM model against the ensemble machine classifier on UCSC motifs \cite{UCSC10}.  Among 88 transcription factors tested the ensemble method performed better for 70 datasets in terms of $F_1$ score (Fig.~\ref{fig:toPWM}).

\begin{figure}[ht] \centering
\includegraphics[width=\textwidth]{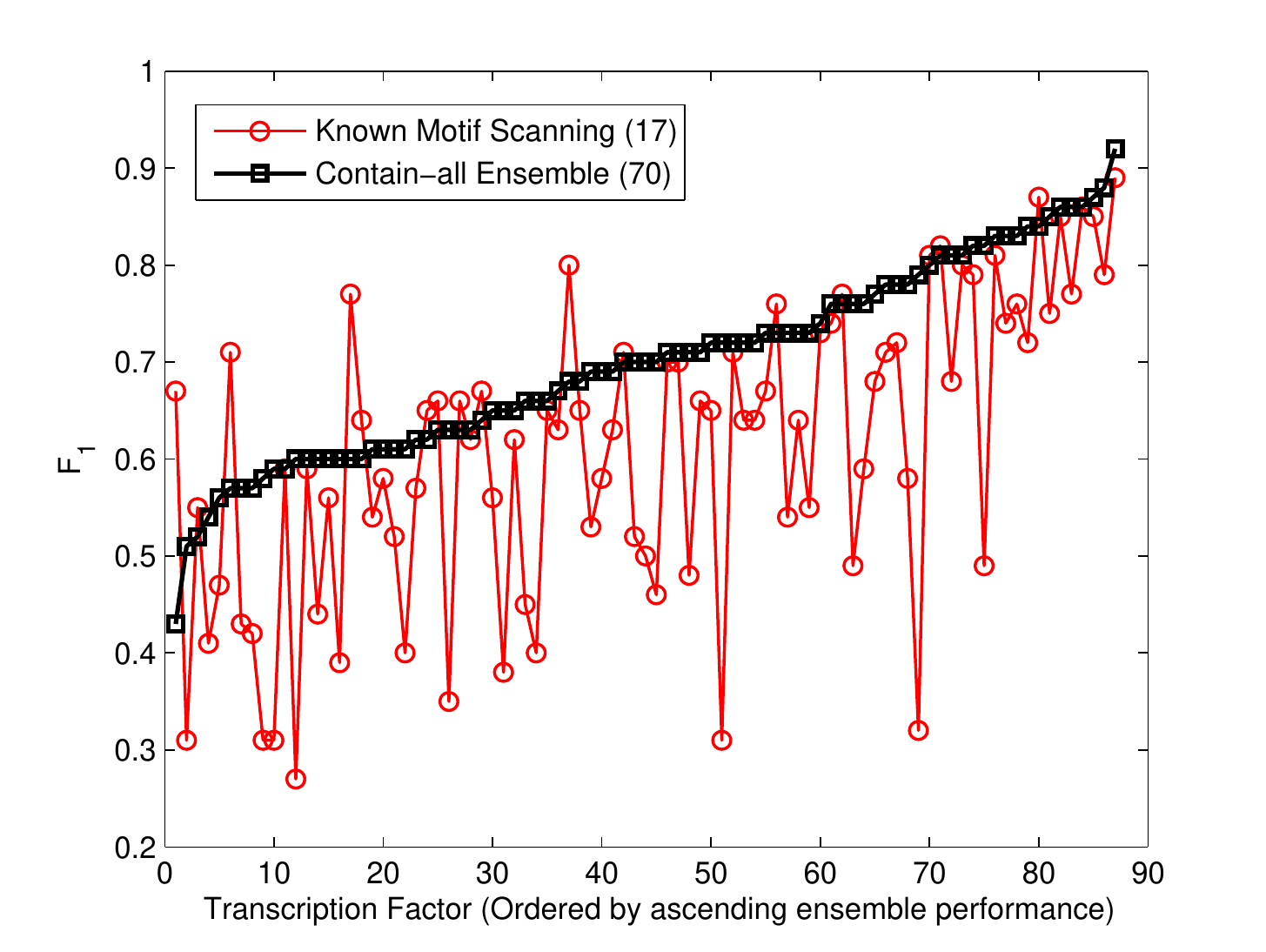} 
\caption{Ensemble classifier performance versus PWM model-based performance on identification of target genes for 88 TFs. $F_1$ scores are based on balanced positive/negative target sets.} 
\label{fig:toPWM}
\end{figure}

\section*{Conclusion}

We have presented an ensemble method for solving the problem of integrating information on transcription factor/gene interactions, and making predictions on this basis.  The approach begins with identifying regulatory target genes and ends with predicting binding locations of TFs. Computational results show the ensemble method can successfully integrate information from five known motif-finding algorithms, each of which focuses on a different optimization problem. The performance is consistently good over variety of data in predicting regulatory target genes and binding site motifs. Generally, machine learning methods have limitations for small sample sizes. The ensemble method together with a useful dimension reduction strategy (using weak learners) gives a tool with high sensitivity (recall) in predicting TF binding sites. The machine learning framework is also an alternative to the PWM model for all three identification problems mentioned earlier.  It improves accuracy in identifying a TF's gene targets and binding motifs, and can also improve binding site identification. The algorithm is scalable, so that more component algorithms and standard motif databases can be added to provide more comprehensive results. There are a number of new TF-DNA databases based on recent experiments \cite{ENCODE} which can contribute significantly to these methods for solving such problems. We have shown that machine classifiers can play a role as an alternative to PWM methods in dealing with the three above-mentioned TF regulation-related problems. With more sophisticated models and methodologies as well as additional data types, this method can scale up and can be of further use.

\section*{Methods}
\label{sec:method}

\subsection*{Overview}
\label{sec:method.0}

For a given TF the computational model we consider is initially based on supervised two-way classification of its target genes (positives) against non-target genes (negatives). This machine learning framework based on $k$-mer feature spaces $F$ was implemented in the development of the SVMotif algorithm \cite{SVMotif}. In SVMotif Ensemble, individual component motif finding algorithms first run through the training dataset $D=\left\{S_i,y_i\right\}_{i=1}^n$. This consists of a given set of promoter sequences $S_i$ containing both positive (known target) and negative (known non-target or presumed non-target) promoter sequences of genes, along with (known) labels $y_i=\pm 1$ for positives/negatives, respectively. For the TF, candidate PWMs are collected using the available algorithms  in the ensemble into an initial candidate pool $\mathcal M$. In practice, candidates could additionally be obtained from standard databases such as Transfac \cite{Transfac} and JASPAR \cite{JASPAR}. 

For each available PWM $M_j \in \mathcal M$, we construct two (independent) classification sub-models. Each takes information from $M_j$, and classifies every novel gene as a target or non-target based on its promoter sequence $S$.  The first sub-model used is the standard PWM scanning model based on $M_j$ (Section \ref{sec:method.2}), while the second is the $k$-mer space SVM model (Section \ref{sec:method.4}). Two kinds of scalar \emph{synopsis features} are extracted from each promoter using the two models: the \emph{rescanning synopsis} and the \emph{subspace synopsis}, denoted as $\phi^r_{M_j}\left(S\right)$ and $\phi^w_{M_j}\left(S\right)$. Each synopsis feature (rescanning and subspace) is a scalar (one dimensional) machine learning (ML) score based on the training samples $\left\{S_i\right\}_{i=1}^N$ and an $M_j$-based sub-classifier.  The sub-classifier is  trained by one of the above two (scanning or subspace) methods to predict the labels $\left\{y_i\right\}$  (see below; recall $y_i=1$ indicates a  target and $y_i=0$ a non-target).  For the TF, the basic gene target finding problem (1 above) and the two additional problems of binding site (2) and motif (3) identification (see Sections \ref{sec:method.6} and \ref{sec:method.7}) are solved using the above classifiers as follows.

\begin{enumerate}
\item For identification of target genes (problem 1) of the TF, an SVM is trained on the ensembles of combined synopsis features. Thus each promoter sequence $S_i$ is mapped into the synopsis feature space $F_s$ via
\[
\phi \left(S_i\right) = \left( \ldots \phi^{r}_{M_j}\left(S_i\right) \ldots  \phi^{w}_{M_j}\left(S_i\right) \ldots  \right)\]
combining classification scores based on all PWM $M_j$ forming both types of scores (the scanning scores $r$ and subspace scores $w$). The $r$ (scanning) scores are based on scoring the promoter sequences $S_i$ using standard scanning with PWM $M_i$ (see below).  On the other hand, the $w$ ($w$-vector) subspace scores are based on scoring the same sequences $S_i$ using the SVM classifier score $f(\mathbf x)$, based on their feature vectors $\mathbf x = \phi(S_i)$ in the $k$-mer feature space $F$.  See Sections \ref{sec:method.1} and \ref{sec:method.4} for a fuller description. The ensemble-based linear (boosting) classifier now trained in the reduced synopsis feature space $F_s$ then has the form of single SVM score 
\[f(S) = \boldsymbol\beta \cdot \phi\left(S\right) +\beta_0\]
used to score out-of-training test sequences $S$.

\item For motif identification (problem 2), a feature selection algorithm is used on the synopsis space $F_s$ to select the most discriminative features (PWMs) in $\phi(S_i)$. Those PWMs corresponding to the top ranked synopsis features (based on discrimination of positive and negative genes) are themselves better PWMs for the true binding motif.  A PWM agglomeration algorithm is then used to collapse similar PWMs into a single motif.
f
\item Finding individual binding sites (problem 3) is then done by both standard PWM-rescanning (based on the PWMs in (2)) and the $w$-rescanning method (see section \ref{sec:method.4}). The union of these two sub-models (based on the $r$ and $w$ scores) for each PWM produces a score for every location on the promoter $S$. These local scores are aggregated through a dot product with the coefficients $\boldsymbol \beta$ learned from the ensemble target classifier (problem 1). The local peaks of this score are identified and reported as predicted binding sites. 

\end{enumerate}

\subsection*{PWM Scanning Models and PWM synopsis features}
\label{sec:method.1}
The PWM model is a widely used motif model in sequence binding analysis \cite{Stormo1990,Stormo2000}.
A PWM corresponding to a TF $t$ is a $4\times k$ matrix, $M=\left( \theta_{ij}
\right)$ whose $j^{th}$ column $\boldsymbol{\theta}_j$ defines the probability
distribution of $\mathcal A=\left\{A,C,G,T\right\}$ appearing at position $j$ in the set of all binding sites of length $k$, i.e., 
\[\theta_{ij} = \mathbf P(i|M,\mbox{Position }j), i\in\mathcal A.\] 
The matrix $M$ is usually generated empirically from a large number of likely binding sites by aligning and then counting frequencies of residues at given positions. In order to identify new potential binding sites within a sequence $S$ (problem 3), the given weight matrix $M$ is first changed into a log-ratio scoring matrix $N$ measuring likelihood against a background distribution $B$, so
\[N_{ij}=\log_2\left(\frac{\theta_{ij}}{b_i}\right),\] 
where $b_i$ is the background probability of observing DNA base
$i \in \mathcal A$. For any subsequence $\alpha=\left(a_1a_2\cdots a_k\right)$ of length $k$ in the promoter, the matrix $N$ can score $\alpha$ using the PWM scanning score
\[s_M^{r}(\alpha)=log\left(\frac{\mathbf P(\alpha|M)}{\mathbf P(\alpha|B)}\right)=\sum_{j=1}^k N_{a_j,j},\]
i.e., is the log-likelihood ratio of observing $\alpha$ under motif model $M$ versus observing it under background $B$. A high score suggests that $\alpha$ is more likely to be a binding site. For a given $M$, PWM scanning scores are computed for all $k$-mers in the promoter sequence. A threshold $\tau$ is then used to identify significant binding $k$-mers.

PWM models have also been used to explore gene targets (problem 1) using certain nonlinear gene candidate scoring systems.  An example is use of the maximal 
local PWM scanning score \textbf{(point 3 at the end of section 2.1 above)} over the entire promoter sequences, with a (different) threshold $\tau$. The explicit form of such a PWM scanning classifier is 
\[f^{r}\left(S\right)= \beta_1 s_M^{r}\left(S\right)-\tau = \beta_1 \max_{\alpha \subset S}s_M^{r}\left(\alpha\right)-\tau,\]
with the maximum over all $k$-mer strings $\alpha$ in $S$, and a value above 0 indicating the gene is a target. Because this maximum is not always robust against noise and outliers, some alternatives based on functions $\rho$ of the $k$-mer scanning scores $s^r(\alpha)$, for $\alpha \subset S$, have been developed to score a promoter $S$.  We can write $s^{r}(S) = \rho \left(\left\{s^{r}(\alpha)\right\}_{\alpha \subset S}\right)$, where $\rho$ can have several forms in addition to the above maximum: 

\begin{enumerate}
\item Linear scoring: $s_M^{r}(S) = \sum_{\alpha \subset S}s_M^{r}(\alpha)$
\item $m$-trimmed linear scoring: $s_M^{r}(S) = \sum_{\mbox{\scriptsize{top~$m$~}}\alpha \subset S}s_M^{r}(\alpha)$
\end{enumerate}
where the latter means the sum is formed not of all rescanning scores, but just the top $m$ scores within the candidate promoter $S$.  Below we have chosen the trimmed linear thresholding score with $m=5$.

A trivial (one dimensional) SVM algorithm can be used on training data consisting of the single feature $s_M^{r}(S)$ to determine the best
$\beta_1$ and $\tau$. Selection of $\tau$ does not change the ranking power
of $s_M^{r}(S)$ or area under the ROC curve of the resulting classifier $f^r$. We
call $s_M^{r}(S)$ the \emph{rescanning synopsis feature} of $S$ corresponding to PWM $M$.


\subsection*{Using the $k$-mer Feature Space and SVM Classifier}
\label{sec:method.2}
String ($k$-mer) representations of sequences have been studied for a number of years \cite{leslie,Dustin2006,SVMotif}.  Any DNA sequence $S$ is an ordered list from the alphabet $\mathcal A=\left\{A,C,G,T\right\}$. Let $\Sigma=\mathcal A^k$ be all DNA strings of length $k$ (denoted as $k$-mers).  Without considering the appearance order or partial overlap of the $k$-mers in the sequence  $S$, we map $S$  into a feature space $F^k$  with feature map 
\[ \phi\left(S\right) = \left\{n_{\alpha_i}\left(S\right)\right\}_{\alpha_i \in  \mathcal A^k},\]
where vector $\phi(S)$ has one component corresponding to each string $\alpha_i$ of length $k$.  Thus the $\alpha^{th}$ element $n_{\alpha}$ of sequence $S$ is a count in this sequence of occurrences of $k$-mer $\alpha$. The feature map $\phi$ is the \emph{spectrum map} \cite{leslie}, as it contains counts of all possible  $k$-mers in $\Sigma$. The feature space $F^k$ of all possible $\phi(S)$ is the $k$-mer (or \textit{spectrum}) feature space. Combining several such spaces with different values of $k$ yields a full string feature space, $F = \oplus F^{(k)}$, where $\oplus$ denotes direct sum over $k$. Thus a feature vector in $F$ is a concatenation of feature vectors in the spaces $F_k$ for allowed $k$ ; without confusion $F$ will sometimes also be called the $k$-mer space.

The full feature map from $S$ into $F$, also denoted as $\phi$, maps training samples $\left\{S_i,y_i\right\}_{i=1}^N$ ($S_i$ is the gene promoter sequence and $y_i = \pm 1$) to the more convenient dataset $D=\left\{\mathbf x_i,y_i\right\}_{i=1}^N$. Here $\mathbf x_i = \phi(S_i) \in F$ is the $k$-mer spectrum of $S_i$. An SVM classifier trained on $D$ can classify new out-of-training samples $S$ \cite{Dustin2006} by computing the discriminant
\[f^{(w)}(S)=\mathbf w \cdot \phi(S) + b_0 = \mathbf w \cdot \mathbf x + b_0.\]
If $f^{(w)}(S) > 0$, the corresponding gene is classified as a binding target. In testing on more than 100 yeast transcription factors, the SVM was trained on a space combining 4, 5 and 6-mer spaces.  The gene classification accuracy was roughly 70\% on test sets balanced between positives and negatives \cite{Dustin2008}.

The $w$-vector $\mathbf w$ of the SVM classifier $f^{(w)}$ has $j^{th}$ component $w_j$ that measures importance of feature $x_i$ (count of the $j^{th}$ $k$-mer) in $S$) \cite{Guyon,RSVM}. Thus $w_j$ measures the power of feature $x_j$ in discriminating  positive ($y=1$) from negative ($y=-1$) training samples. Note here the index $j$ and index $\alpha$ are considered the same.  Table~\ref{tab:topkmers} shows the level of association between top ranking $k$-mers and known binding motifs of TF's. Based  on  this,  the SVMotif algorithm \cite{SVMotif} was developed to extract binding motifs from  the $w$-vector by agglomerating $k$-mers $\alpha$ with top feature importance scores $w_\alpha$ \cite{SVMotif}. Tested on 85 yeast TFs with known binding motifs \cite{Harbison04, MacIsaac2006}, SVMotif was able to correctly predict 40 standard motifs in its top prediction and 57 of them within its top 3 predictions \cite{SVMotif}. 

\begin{table}
\footnotesize
\centering
\begin{tabular}{c|p{.6cm}p{.6cm}|p{.6cm}p{.6cm}|p{.6cm}p{.6cm}|p{.6cm}p{.6cm}}
\hline\hline
Rank&\multicolumn{2}{c|}{GCN4}&\multicolumn{2}{c|}{UME6}&\multicolumn{2}{c|}{MIG1}&\multicolumn{2}{c}{STE12}\\
\hline
1&\textbf{gagtca}  &8.099&\textbf{gccgcc}  &5.262&\textbf{cccgc}   &0.858&\textbf{gtttca}  &2.764\\
2&\textbf{agtca}t  &4.434&\textbf{agccgc}  &5.103&\textbf{gggg}aa  &0.658&cgagaa  &1.084\\
3&\textbf{gactca}  &4.094&\textbf{cggcta}  &4.253&a\textbf{cccca}  &0.622&\textbf{gaaaca}  &1.014\\
4&\textbf{agtca}   &1.679&\textbf{gccgc}   &3.04&\textbf{ccccgc}  &0.616&cattcc  &0.934\\
5&\textbf{gactc}   &1.66&\textbf{ccgcc}    &2.718&\textbf{ccgga}   &0.604&tcctaa  &0.79\\
6&\textbf{agtca}c  &1.127&\textbf{ccgccg}  &2.05&a\textbf{cccc}   &0.603&agtatg  &0.708\\
7&cattag  &0.933   &\textbf{cgccga}        &1.857&\textbf{ccgg}    &0.597&acattc  &0.649\\
8&cttatc  &0.886   &\textbf{agccg}         &1.634&\textbf{ccccac}  &0.568&\textbf{aaaca}g  &0.609\\
9&\textbf{actca}   &0.832&c\textbf{gccg}   &1.124&ccgta   &0.556&a\textbf{tgaaa}  &0.562\\
10&ca\textbf{tgac} &0.734&g\textbf{cgcc}   &0.845&gcaaca  &0.524&taggaa  &0.556\\
\hline\hline
\end{tabular}
\hspace{\baselineskip}
\caption{Top 10 $k$-mers in the output for sample yeast transcription factors
GCN4(TGACTCA), UME6(TAGCCGCCSA), MIG1(WWWWSYGGGG), and STE12(TGAAACA). 
True (standard) binding motifs, retrieved from 
YeastGenome [www.yeastgenome.org], are listed in UPPER case following the gene name above. $k$-mers 
matching the corresponding motifs are highlighted in \textbf{bold}. 
The feature importance scores $I_j$ for $k$-mer features are listed 
next to the corresponding $k$-mers.}
\label{tab:topkmers}
\end{table}

We note, as briefly mentioned earlier, that this SVM classifier can be viewed as a generalization of the standard PWM re-scanning classifier with linear scanning score (section \ref{sec:method.1}). Specifically, if each $w$-component $w_j$ were to be artificially constructed to equal the PWM scanning score of the corresponding $k$-mer $\alpha_j$ (so $w_j=s_M^r(\alpha_j)$), then the linear motif rescanning classifier could be written as
\[f^{r}(S) =  \sum_{\alpha
\subset S} s_M^{r}(\alpha) - \tau = \sum_j s^{r}(\alpha_j)n_{\alpha_j} -  \tau
= \sum_j w_j x_j -  \tau.\] 
Here the first sum is over all substrings $\alpha$ of $S$ of length $k$, while the second is over all possible $k$-mers $\alpha_j \in \Sigma$.  Each element $w_{Mj}$ of $\mathbf w_M$ is the linear PWM scanning score of $M$ on $k$-mer $\alpha_j$. From this viewpoint any PWM  $M$ corresponds to a $w$-vector, which we denote as $\mathbf w_M$, such that the SVM-based classifier  $f=\mathbf w_M  \cdot \mathbf  x +  b$ is  equivalent to the PWM scanning classifier function with linear score. We call such vectors $w=w_M$ \emph{PWM-compatible $w$-vectors}.

On the other hand, in standard SVM training on a sample set $D$, the optimal SVM $w$-vector is obtained from optimizing (minimizing) the Lagrangian 
\[ \mathcal L = C\cdot \sum_i
Loss\left(y_i,   f(\mathbf  x_i)\right)   +  \frac{1}{2}\left\|w\right\|^2.  \]
Here $Loss\left(y_i,f(\mathbf x_i)\right) = \left(1-y_if(\mathbf x_i)\right)_+$ is the standard SVM hinge loss function; recall that 

\begin{eqnarray}
(a)_+ = \left\{ \begin{array}{rl}
	              a & \mbox{  if $a>=0$, } \\
	      0 & \mbox{  if $a<0$.  }
       \end{array} \right. \nonumber
\end{eqnarray}

Since this optimization searches the entire  $w$-vector space (including all PWM-compatible $w$-vectors) this optimized SVM classifier will be systematically as good as or better than (by the optimization criterion) any PWM-based scanning classifier with linear thresholding function.

However, not all PWM re-scanning classifiers are based on linear scoring.  
The classifier with maximum scanning score 
\[f^{r}(S) = \max_{\alpha \subset S} s_M^{r}(\alpha) - \tau, \]
defined earlier in Section \ref{sec:method.1}, can similarly be emulated in this ML framework using another sub-class of SVM-based classifiers, as shown below.

This time we replace $F$ by a modified feature space $F'$ in which each component $x'_j$ of feature vector $\mathbf x'$ is binary, i.e. either value 1 or 0.  This  depends on whether $k$-mer $\alpha_j$ does or does not appear in the promoter $S$.  In practice it has been shown \cite{SVMotif} that such a modified feature space $F$ can improve accuracy of motif identification.  We begin with the mathmetical fact that the $L^p$ norm approaches the $L^\infty$ norm as $p\rightarrow\infty$, i.e.,  for any numbers $v_j$, 
\[ \left(\sum_j v_j^p\right)^\frac{1}{p} \rightarrow \max_j|v_j| \mbox{ as } p \rightarrow \infty.\]
Based on this we can approximate the maximum scanning score (with large $p$) by  
\[\max_j(w_jx'_j) \approx \left(\sum_j w_j^p x'_j\right)^\frac{1}{p} .\]
Here $w_j^p$ is the $p^{th}$ power of the PWM scanning score $w_j$ of $k$-mer $\alpha_j$, and $x'_j = 1$ or $0$ indicates whether $n_{\alpha_j}>0 or not$. For very large $p$, the term $\left(\sum_j w_j^p x'_j\right)^\frac{1}{p}$ approximates the maximum scanning score of $S$, i.e., 
\[ \max_j(w_jx'_j) = \max_{j:x'_j=1}w_j=\max_{\alpha \subset S} s_M^{r}(\alpha).\]; recall that $s_M^r$ denotes the (re)-scanning score with PWM $M$ of $k$-mer $\alpha$.

The monotonic increase in $x$ of the function $f(x)=x^{1/p}$, ensures that the SVM scoring function $\sum_j w_j^p x'_j$ always ranks the 
samples in the same way as $\left(\sum_j w_j^p x'_j\right)^\frac{1}{p}$. Thus the approximation 
\[f_1^{r}(S) = \sum_j w_j^p x'_j + b, \]
with choice of threshold $b=-\tau^p$ can be constructed to select the same targets as the original PWM-based thresholding classifier $f^{r}$ based on the maximum scanning score.

Note  that  the new  $w$-vector  $\mathbf w'$,  with ($w'_j=w_j^p$), is still in the search space of the SVM optimization. Therefore after finding the $\mathbf w$ minimizing the above Lagrangian $\mathcal L$ over the space of {\it all} candidate $\mathbf w$'s (including the above PWM-based $w$-vectors), the quality of the resulting SVM (as defined by lower values of $\mathcal L$ itself) is also as good as or better than the maximum score PWM-based classifier.

\subsection*{PWM subspaces}
\label{sec:method.3}
In the PWM rescanning model, a gene is classified as a TF target if its promoter contains $k$-mers that are high-scoring with respect to to the PWM motif of the TF.  The frequency of $k$-mers with high PWM scores in a promoter is strongly correlated with the PWM-based classifier predicting the promoter as a target. Thus in the $k$-mer feature space $F$, an ML algorithm based on $F$ can be dimensionally reduced (without much loss) to operate within the subspace $F_M$ that is spanned only by basis $k$-mers that are high-scoring with respect to $M$.

For a given $M$, we define the set of such high-scoring $k$-mers to be the \emph{profile set}, denoted as $\Sigma_M$.  We call the feature subspace $F_M$ the \emph{profile subspace} of $F$ corresponding to $M$. A motif discovery algorithm producing such dimensionally reduced subspaces of the full $F$ (e.g., based as above on a PWM) is then called a \emph{subspace-valued weak learner}.  For a single $M$, the ML term `weak learner' refers to the fact that the subspace $F_M$ learned from $M$ is a small part of a sum of such subspaces aggregated from different candidate $M$ for the TF (it does not necessarily denote weakness in its ususal sense).  Unlike standard weak learners, which make individual predictions that are then aggregated, these weak learners produce dimensionally reduced candidate subspaces for further ML search.  This is based on the ML notion that individual weak learners can be aggregated into stronger machines.

Given PWM $M$, the $k$-mers $\alpha$ in the profile set $\Sigma_M$ spanning the subspace $F_M$ are selected as follows.
Suppose $M=(\theta_{ij})$ is a $4 \times l$ PWM. From $M$ we can randomly generate length $l$ DNA base sequences
$\alpha=\left(a_1 a_2\ldots a_l\right)$ by selecting the letters $a_j$ 
from a $multinomial\left(\mathbf{\theta}_{\cdot j}\right)$ probability distribution, 
with $\mathbf{\theta}_{\cdot j}$ the $j^{th}$ column vector of $M$, giving the probability distribution
of bases ($A,C,G$ or $T$) at position $j$. To generate
shorter sequences ($k<l$), a smaller block $M'$, 
which consists of $k$ consecutive columns of $M$, is randomly selected and used to replace $M$. 
On the other hand, to generate a longer sequences ($k>l$), columns generated by a background nucleotide distribution can be concatenated to expand $M$ on both sides.

After generating a representative set of $k$-mers from $M$ in this way, 
an additional step is to score each $k$-mer using the log-ratio scoring matrix $N$ 
above and to eliminate those that score below a set threshold. This ensures that
the remaining $k$-mers follow a pattern significantly different 
from the background. Among many different ways to choose the threshold, we used 
\[\tau = 0.85* \max_{\alpha \in \Sigma}{s^{r}_{M}\left(\alpha\right)}.\]
 \textbf{[we need to explain the choice - is it the one that worked best overall?]}

For $k=6$, this method results in a $6$-mer subspace of dimension 50 to 200, out of an original dimensionality of 2080 \footnote{$k$-mers forming reverse complements are treated as the same feature, so the dimension is smaller than  the standard $4^6$}. In our implementation we consider $k$-mers of length from 4 to 10. If a PWM $M$ succeeds in capturing the pattern of binding sites of the TF, the $k$-mers in $\Sigma_M$ should be sufficient to separate binding target sequences from non-targets. Thus the SVM built in the dimensionally reduced $F_M$ should perform similarly to that built in the full feature space $F$, or better if noise reduction is taken into account. From this point of view the component algorithms providing $M$'s can be seen as dimension reduction tools that filter out noisy $k$-mers from the feature space. Because the dimension is significantly reduced, machine learning methods can now also be used to examine longer $k$-mer patterns. 

We give a numerical example to illustrate the connections among the above-mentioned types of classifiers. These include PWM rescanning classifiers (using maximum  scanning scores),  the SVM classifiers 
\[f(S)=\sum_j w_j^{p}x'_j+b, \mbox{ with } w_j=s^r(\alpha_j) \mbox{ and } p=1,4,10,\] 
linear SVM classifiers (trained on full $k$-mer feature space $F$), and subspace-based SVM classifiers (trained on the $M_0$ profile subspace;  see \ref{sec:method.2} and \ref{sec:method.3}). The PWM $M_0$ for a well studied yeast TF, GCN4, was retrieved  from the UCSC database \cite{UCSC10},  and positive  and negative  samples of  target and non-target  promoters  were  obtained   (based  on  ChIP-chip  experiment data \cite{Lee_Network02}).

Note that the above PWM-based rescanning classifier and the SVM-based classifiers with a PWM-compatible $w$-vector do not involve training (after selection of the PWM), since the $w$-vector is obtained strictly using PWM-based $k$-mer scanning scores, i.e., $w_j=s^{r}_{M_0}\left(\alpha_j\right)$ (with $\alpha_j$ the $k$-mer corresponding to the component $w_j$ of $\mathbf w$). Thus they are tested on the whole dataset directly. The two SVM classifiers in table \ref{tab:ensemble.machine.roc} were assessed  under a  5-fold cross-validation  protocol, where all samples are randomly divided into 5 portions, with the classifier trained on 4 portions and scored (tested) on the remaining portion - this is repeated 5 times, rotating the test portion, until all samples are scored. Table  \ref{tab:ensemble.machine.roc}  shows  the  area  under  the  ROC  curve  for each algorithm, and shows that for sufficiently large $p$ the SVM-based classifier has about the  same  performance as  the  PWM-scanning classifier. This is consistent with the argument that the search space of the second classifier is in fact contained in the other (Section \ref{sec:method.2}).  Interestingly, the dimension-reduced  SVM classifier trained on the subspace $F_{M_0}$ (the subspace of $k$-mers generated by $M_0$, here restricted to 35 dimensions)  outperforms the  SVM classifier  built on  the full  space ($4^7$  dimensions),  suggesting the  effectiveness  of  dimension reduction through PWM-subspaces (Section \ref{sec:method.3}). 

\begin{table}
\footnotesize
\centering
\label{tab:ensemble.machine.roc}
\begin{tabular}{cccccc}
\hline\hline
PWM&SVM-based(1)&SVM-based(4)&SVM-based(10)&SVM ($F$)&SVM ($F_{M_0}$)\\
\hline
0.8428&0.6691&0.8282&0.8441&0.6833&0.8195\\
\hline\hline
\end{tabular}
\hspace{\baselineskip}
\hspace{\baselineskip}
\hspace{\baselineskip}
\caption{Area under ROC  curves (AUROC) for different classifiers. The PWM column presents the AUROC of the PWM maximum scanning score-based classifier. Columns 2 through 4 present the AUROC of the above-mentioned SVM classifier using the $p^{th}$ powers of PWM-compatible $w$-vectors ($p=1,4,10$). The two columns on the right present the AUROC for two linear SVMs, one ($F$) trained on full 7-mer feature space and one ($F_{M_0}$) on the $M_0$-derived 7-mer subspace. }
\end{table}

\subsection*{Subspace Synopsis Features and the $w$-Scanning Model}
\label{sec:method.4}

The profile set and profile subspace corresponding to a PWM $M$, defined the previous section, are used to reduce dimension, focusing classifiers on the parts of the feature space $F$ with the most information.  Here we will define some tools arising out of profile subspaces.   First the profile set $\Sigma_M$ of a good PWM $M$ has $k$-mers whose frequency counts in a promoter can discriminate positives (targets) and negatives (non-targets) easily. Thus the discrimination power of the SVM classifier trained just in subspace $F_M$, given as 
\[
f_M(S)=\sum_{j:\alpha_j \in \Sigma_M } w_jx_j + b,
\]
also provides a goodness measure of the corresponding $M$. 

Second, because the shift $b$ does not affect gene rankings, we will use scores without $b$, given by
\[ s_M^{w}(S) = \sum_{j:\alpha_j \in \Sigma_M } w_jx_j \],
as a subspace synopsis feature of the classifier built from $M$.  A single synopsis feature such as this summarizes information from the subspace $F_M$, while the union of such features (over different $M$ from the ensemble) forms the feature vector for discriminating the TF's targets.

Third, for binding site detection we use a $w$-vector scanning model to find the binding sites in a promoter $S$. Instead of using the standard PWM scanning-based log-ratio scoring matrix $N$ (Section \ref{sec:method.1}) to score each $k$-mer string $\alpha$ in $S$ (as a potential binding site), the $w$-scanning model provides $\alpha$ with an SVM feature importance score \cite{RSVM},
\[s^w_M(\alpha)=w_\alpha\left(\bar{x}_\alpha^{(+)}-\bar{x}_\alpha^{(-)}\right).\]
Here the $w_\alpha$ are coefficients in the trained classification function $f_M(g)$, and $\bar{x}_\alpha^{(+)}$ and $\bar{x}_\alpha^{(-)}$ are average numbers of appearances of $\alpha$ in the positive and negative training sets respectively. Scanning along the promoter $S$, successive feature importance scores $\left(s^w_M(\alpha_1),s^w_M(\alpha_2),\ldots,s^w_M(\alpha_n)\right)$ give for $k$-mer $\alpha_j$ at position $j$ an importance score $s^w_M(\alpha_j)$. As is the case for the PWM-based rescanning, a threshold is determined for selecting $k$-mers $\alpha_j$ giving significant predictions (Section \ref{sec:method.7}).

\subsection*{Identifying Gene Targets with Ensembles Using Synopsis Feature Spaces}
\label{sec:method.5}
With the basics above we construct in more detail the ensemble methods for our identification problems.  We use five component ensemble algorithms, or weak learners, each generating PWMs to be used in the ensemble algorithm. 

For each TF we trained these five components on the training data sequences (known positives and negatives), without any parameter tuning. For BioProspector and MEME, the size of the motif is a required parmeter, and we therefore used multiple runs of
each algorithm with motif sizes from 7 to 12.  For each TF we collected 29 PWMs (see Section \ref{sec:result.1}) from the five algorithms into a candidate pool, 
$\mathcal{M}=\{M_1,\ldots,M_{29}\}$. 
With these PWMs we constructed the (synopsis) feature spaces defined below for their ensemble estimates of the best PWM.
  
\textbf{Rescanning Features}: For each PWM $M_j \in \mathcal{M}$, we used the PWM rescanning synopsis scores $s^{r}_{M_j}(S)$ as features mapping both training and test data into the rescanning synopsis feature space $F^r_s$ with feature vectors 
\[ \phi^{r}(S) = \left(s^{r}_{M_1}(S),s^{r}_{M_2}(S),\ldots,s^{r}_{M_{29}}(S)\right) \]

\textbf{Subspace Features}: In addition to  using PWM rescanning synopsis scores as features, we also studied the subspace-based synopsis features. We first trained a dimensionally reduced SVM in the (previously mentioned) subspaces $F_{M_j}$ determined by PWMs $M_j$. Then each sequence $S$ (in the training or test data) was mapped into the subspace synopsis feature space $F^w_s$ with feature vectors
\[ \phi^{w}(S) = \left(s^{w}_{M_1}(S),s^{w}_{M_2}(S),\ldots,s^{w}_{M_{29}}(S)\right). \]
This formed a second set of features for discriminating genes as targets or non-targets of the TF.

\textbf{Ensemble Features}: 
We combined the features from the rescanning and subspace synopsis to get 
the full synopsis feature vector \[\phi^{c}(S)=\left(\phi^{r}(S),\phi^{w}(S)\right). \]
The resulting full (but still low dimensional) synopsis feature space $F_s=F^r_s\oplus F^w_s$ thus has 58 dimensions. This space allows for using standard low dimensional statistical methods to classify targets of the TF based on these features.

We can now use the above feature spaces to solve problem (1), the discrimination of gene targets of the TF. Our feature maps send training and test samples into synopsis spaces $F^r_s$, $F^w_s$ or (for the combined map) $F_s$. An SVM is built in each of these spaces to form an ensemble-based classifier determining the TF targets and non-targets, from the original weak classifiers $s^r_{M_j}(S)$ or $s^w_{M_j}(S)$. Thus the full ensemble SVM discriminant function (separating targets and non-targets) is $f_s^{c}(S)=\boldsymbol \beta \cdot \phi^{c}(S) + \beta_0$. 

Below we describe the algorithms for identifying binding motifs and binding sites based on this ensemble. For definiteness we assume the sample sequence $S$ is mapped into the synopsis feature space $F_s = F^r_s\oplus F^w_s$ using the above feature map $\phi^c$. 

\subsection*{Identifying Binding Motifs with Ensembles}
\label{sec:method.6}
 
Our ML approach formalizes the problem of finding binding motifs for a given TF as a feature selection problem. Specifically, sample sequence $S$ (among the known targets or non-targets) is represented in the combined synopsis space $F_s$ by feature vector
\[ \phi^{c}(S) = \left(s^{r}_{M_1}(S),\ldots,s^{r}_{M_{29}}(S),s^{w}_{M_1}(S),\ldots,s^{w}_{M_{29}}(S)\right). \]
Selection of the best PWM $M_j$ will mean finding the one giving the most discriminative feature in the feature set $\phi^c(S)$.  Here discrimination is measured based on target/non-target separation in the training set of known targets (positives)/non-targets (negatives).  Thus we select the set of the above PWMs whose synopsis features (rescanning synopsis $s^{r}_{M_j}(S)$ or subspace synopsis $s^{w}_{M_j}(S)$) jointly form the best set for discriminating targets.

Among a number of feature selection methods for SVM, we tested RSVM \cite{RSVM} within $F_s$. In each run the machine generates an importance score for each synopsis feature ($s^{r}_{M_j}(S)$ or $s^{w}_{M_j}(S)$) from which we rank the $M_j \in \mathcal{M}$. 

We repeated the feature selection procedure multiple times with different sets of negatives (presumed non-target sequences selected at random from the genome) together with the known set of positives. The averaged RSVM importance score was finally used to rank PWMs in $\mathcal{M}$. Because some PWMs in $\mathcal M$ are highly similar to each other, we also clustered the 10 top PWMs and re-ranked them using their weighted entropy scores (see \cite{SVMotif} for the score definition).

The reported PWMs were compared with UCSC motifs \cite{UCSC10} using the following standard motif similarity measure. Because each PWM column defines a distribution among $\mathcal A = \{A,C,G,T\}$, we first define the similarity between two PWM columns $P$ and $Q$ as based on Jensen-Shannon divergence (symmetrized Kullback-Leibler divergence), defined as the distance 
\[ \rm{Sim}\left(P, Q\right)=\frac{1}{2}D_{KL}\left(P\|Q\right)
	+\frac{1}{2}D_{KL}\left(Q\|P\right)\]
between the two probability distributions, where $D_{KL}\left(P\|Q\right)=\sum_i 
P_{i}\log\left(P_{i}/Q_{i}\right)$.
Here $P_i$ and $Q_i$ represent probabilities of points $i$ in the probability space, corresponding to distributions $P$ and $Q$.  Then given an alignment $\pi$ (giving a correspondence) between the columns of two PWMs $M$ and $M_0$, the similarity is the sum of similarity scores between pairs of aligned columns $M_j$ and $M_{0j}$, i.e., 
\[
\rm{Sim}_\pi\left(M, M_0\right) = \sum_j Sim\left(M_j, M_{0j}\right).\]
The final similarity is the maximum among all possible non-gapped alignments,
\[ \rm{Sim}\left(M, M_0\right) = \max_\pi{Sim_\pi\left(M, M_0\right)}
\]

\subsection*{Identifying Binding Sites with Ensembles}
\label{sec:method.7}
We have discussed the PWM scanning model (Section \ref{sec:method.1}) and the $w$-scanning model (Section \ref{sec:method.4}) for identifying binding sites. Either model can generate a series of binding strength scores at the local $k$-mer level, i.e., for each $k$-mer within promoter $S$. Each consecutive $k$-mer $\alpha_j$ in $S$ yields a local feature vector 
\[
\mathbf s_j=\left(s^r_{M_1}(\alpha_j),\ldots,s^r_{M_{29}}(\alpha_j),s^w_{M_1}(\alpha_j),\ldots,s^w_{M_{29}}(\alpha_j)\right)
\]
of 58 scores, 29 PWM scanning scores and 29  $w$-scanning scores. Comparing with the global definition of $\phi^c(S)$ for the entire promoter, this vector is a local version, applied to $k$-mers in $S$. The local ensemble score of the $k$-mer is defined as a linear combination of these 58 scores with the same coefficients $\boldsymbol \beta$ as the ensemble SVM classifier for finding gene targets (Section \ref{sec:method.5}).  Thus coefficients $\beta_i$ of the classifier $f_s^c(S)=\boldsymbol \beta \cdot \phi^c(S)+\beta_0$ for gene target identification are the components of $\boldsymbol \beta$, now used locally with the above feature vectors $\mathbf s_j$ to score consecutive $k$-mers $\alpha_j$ as candidate binding sites. The local ensemble score for each $\alpha_j$ is thus computed as
\[ 
s^c(\alpha_j)=\boldsymbol \beta \cdot \mathbf s_j.
\]

We use the same scheme to score locally both positives and negatives. The scores for negatives can provide a background distribution for this class of comprehensive local scanning scores. Then the scores of positives are standardized by subtracting the mean of the background scores and dividing by their standard deviation.  The threshold 2.575 has the property that 10\% of the negatives score above it, thus yielding a 10\% false positive rate.  The $k$-mer locations in the promoter $S$ yielding local scoring peaks above 2.575 ($s(\alpha_j)>2.575$) are identified as potential binding sites at the first stage of the algorithm.

These raw potential sites usually self-aggregate into several motif patterns. Therefore, in order to further remove false positives, we cluster the motif patterns with a greedy algorithm \cite{SVMotif}. The potential binding sites matched to the primary cluster are finally reported as second stage predicted binding sites.  

Because the algorithm used in the second (clustering) stage is greedy, some $k$-mers that temporarily match the pattern of the primary cluster may not match it at the end of the clustering process. We then use the following method to select the final (third stage) matched sites. We scan each potential binding site $\sigma$ from the first stage with the PWM $M_p$ of the primary cluster from the second stage. A normalized rescanning score $NS(\sigma)$ is then calculated as
\[ 
NS(\sigma) = (s_{M_p}^{r}(\sigma)-MIN)/(MAX-MIN),
\] 
where $s^r_{M_p}(\sigma)$ is the rescanning score of the $M_p$ (see section \ref{sec:method.1}) and MAX/MIN are best and worst rescanning scores obtained by scanning all possible substrings with 
the PWM $M_p$. If $NS (\sigma)> 0.8$, then $\sigma$ is kept in the final (third stage) prediction.

\bigskip


 


\newcommand{\BMCxmlcomment}[1]{}

\BMCxmlcomment{

<refgrp>

<bibl id="B1">
  <title><p>{What is a gene, post-ENCODE? History and updated
  definition}</p></title>
  <aug>
    <au><snm>Gerstein</snm><fnm>M. B.</fnm></au>
    <au><snm>Bruce</snm><fnm>C.</fnm></au>
    <au><snm>Rozowsky</snm><fnm>J. S.</fnm></au>
    <au><snm>Zheng</snm><fnm>D.</fnm></au>
    <au><snm>Du</snm><fnm>J.</fnm></au>
    <au><snm>Korbel</snm><fnm>J. O.</fnm></au>
    <au><snm>Emanuelsson</snm><fnm>O.</fnm></au>
    <au><snm>Zhang</snm><fnm>Z. D.</fnm></au>
    <au><snm>Weissman</snm><fnm>S.</fnm></au>
    <au><snm>Snyder</snm><fnm>M.</fnm></au>
  </aug>
  <source>Genome Research</source>
  <publisher>Program in Computational Biology \& Bioinformatics, Yale
  University, New Haven, Connecticut 06511, USA;</publisher>
  <pubdate>2007</pubdate>
  <volume>17</volume>
  <issue>6</issue>
  <fpage>669</fpage>
  <lpage>-681</lpage>
</bibl>

<bibl id="B2">
  <title><p>{Regulatory gene networks and the properties of the developmental
  process}</p></title>
  <aug>
    <au><snm>Davidson</snm><fnm>EH</fnm></au>
    <au><snm>McClay</snm><fnm>DR</fnm></au>
    <au><snm>Hood</snm><fnm>L</fnm></au>
  </aug>
  <source>Proceedings of the National Academy of Sciences of the United States
  of America</source>
  <publisher>Division of Biology, California Institute of Technology, Pasadena,
  CA 91125, USA. davidson@caltech.edu</publisher>
  <pubdate>2003</pubdate>
  <volume>100</volume>
  <issue>4</issue>
  <fpage>1475</fpage>
  <lpage>-1480</lpage>
</bibl>

<bibl id="B3">
  <title><p>{Transcription regulation and animal diversity}</p></title>
  <aug>
    <au><snm>Levine</snm><fnm>M</fnm></au>
    <au><snm>Tjian</snm><fnm>R</fnm></au>
  </aug>
  <source>Nature</source>
  <publisher>Department of Molecular and Cell Biology, Division of Genetics and
  Development, Center for Integrative Genomics, University of California,
  Berkeley, 401 Barker Hall, Berkeley, California 94720, USA.
  mlevine@uclink4.berkeley.edu: Nature Publishing Group</publisher>
  <pubdate>2003</pubdate>
  <volume>424</volume>
  <issue>6945</issue>
  <fpage>147</fpage>
  <lpage>-151</lpage>
</bibl>

<bibl id="B4">
  <title><p>{Systematic identification of mammalian regulatory motifs' target
  genes and functions}</p></title>
  <aug>
    <au><snm>Warner</snm><fnm>JB</fnm></au>
    <au><snm>Philippakis</snm><fnm>AA</fnm></au>
    <au><snm>Jaeger</snm><fnm>SA</fnm></au>
    <au><snm>He</snm><fnm>FS</fnm></au>
    <au><snm>Lin</snm><fnm>J</fnm></au>
    <au><snm>Bulyk</snm><fnm>ML</fnm></au>
  </aug>
  <source>Nature Methods</source>
  <publisher>Nature Publishing Group</publisher>
  <pubdate>2008</pubdate>
  <volume>5</volume>
  <issue>4</issue>
  <fpage>347</fpage>
  <lpage>-353</lpage>
</bibl>

<bibl id="B5">
  <title><p>{A survey of DNA motif finding algorithms}</p></title>
  <aug>
    <au><snm>Das</snm><fnm>M</fnm></au>
    <au><snm>Dai</snm><fnm>HK</fnm></au>
  </aug>
  <source>BMC Bioinformatics</source>
  <pubdate>2007</pubdate>
  <volume>8</volume>
  <issue>Suppl 7</issue>
  <fpage>S21+</fpage>
</bibl>

<bibl id="B6">
  <title><p>Identification of consensus patterns in unaligned DNA sequences
  known to be functionally related</p></title>
  <aug>
    <au><snm>Hertz</snm><fnm>GZ</fnm></au>
    <au><snm>Hartzell</snm><fnm>GW</fnm></au>
    <au><snm>Stormo</snm><fnm>GD</fnm></au>
  </aug>
  <source>Computational and Applied Biosciences</source>
  <pubdate>1990</pubdate>
  <volume>6</volume>
  <fpage>81</fpage>
  <lpage>92</lpage>
</bibl>

<bibl id="B7">
  <title><p>{ANN-Spec: a method for discovering transcription factor binding
  sites with improved specificity.}</p></title>
  <aug>
    <au><snm>Workman</snm><fnm>C. T.</fnm></au>
    <au><snm>Stormo</snm><fnm>G. D.</fnm></au>
  </aug>
  <source>Pacific Symposium on Biocomputing</source>
  <publisher>Center for Biological Sequence Analysis, Technical University of
  Denmark, Lyngby, Denmark. workman@cbs.dtu.dk</publisher>
  <pubdate>2000</pubdate>
  <fpage>467</fpage>
  <lpage>-478</lpage>
</bibl>

<bibl id="B8">
  <title><p>Finding DNA regulatory motifs within unaligned noncoding sequences
  clustered by whole-genome mRNA quantitation</p></title>
  <aug>
    <au><snm>Roth</snm><fnm>FP</fnm></au>
    <au><snm>Hughes</snm><fnm>JD</fnm></au>
    <au><snm>Estep</snm><fnm>PW</fnm></au>
    <au><snm>Church</snm><fnm>GM</fnm></au>
  </aug>
  <source>Nature Biotechnology</source>
  <pubdate>1998</pubdate>
  <volume>16</volume>
  <fpage>939</fpage>
  <lpage>945</lpage>
</bibl>

<bibl id="B9">
  <title><p>BioProspector: discovering conserved DNA motifs in upstream
  regulatory regions of co-expressed genes</p></title>
  <aug>
    <au><snm>Liu</snm><fnm>X</fnm></au>
    <au><snm>Brutlag</snm><fnm>DL</fnm></au>
    <au><snm>Liu</snm><fnm>JS</fnm></au>
  </aug>
  <source>Proceedings of the Sixth Pacific Symposium on Biocomputing</source>
  <pubdate>2001</pubdate>
  <fpage>127</fpage>
  <lpage>138</lpage>
</bibl>

<bibl id="B10">
  <title><p>Unsupervised learning of multiple motifs in biopolymers using
  expectation maximization</p></title>
  <aug>
    <au><snm>Bailey</snm><fnm>TL</fnm></au>
    <au><snm>Elkan</snm><fnm>C</fnm></au>
  </aug>
  <source>Machine Learning</source>
  <pubdate>1995</pubdate>
  <volume>21</volume>
  <fpage>51</fpage>
  <lpage>80</lpage>
</bibl>

<bibl id="B11">
  <title><p>{MEME: discovering and analyzing DNA and protein sequence
  motifs}</p></title>
  <aug>
    <au><snm>Bailey</snm><fnm>TL</fnm></au>
    <au><snm>Williams</snm><fnm>N</fnm></au>
    <au><snm>Misleh</snm><fnm>C</fnm></au>
    <au><snm>Li</snm><fnm>WW</fnm></au>
  </aug>
  <source>Nucleic Acids Research</source>
  <pubdate>2006</pubdate>
  <volume>34</volume>
  <issue>suppl\_2</issue>
  <fpage>W369</fpage>
  <lpage>373</lpage>
</bibl>

<bibl id="B12">
  <title><p>An algorithm for finding protein-DNA binding sites with
  applications to chromatin-immunoprecipitation microarray
  experiments</p></title>
  <aug>
    <au><snm>Liu</snm><fnm>XS</fnm></au>
    <au><snm>Brutlag</snm><fnm>DL</fnm></au>
    <au><snm>Liu</snm><fnm>JS</fnm></au>
  </aug>
  <source>Nature Biotechnology</source>
  <pubdate>2002</pubdate>
  <volume>20</volume>
  <fpage>835</fpage>
  <lpage>-839</lpage>
</bibl>

<bibl id="B13">
  <title><p>Detecting subtle sequence signals: a Gibbs sampling strategy for
  multiple alignment</p></title>
  <aug>
    <au><snm>Lawrence</snm><fnm>CE</fnm></au>
    <au><snm>Altschul</snm><fnm>SF</fnm></au>
    <au><snm>Boguski</snm><fnm>MS</fnm></au>
    <au><snm>Liu</snm><fnm>JS</fnm></au>
    <au><snm>Neuwald</snm><fnm>AF</fnm></au>
    <au><snm>Wootton</snm><fnm>JC</fnm></au>
  </aug>
  <source>Science</source>
  <pubdate>1993</pubdate>
  <volume>262</volume>
  <fpage>208</fpage>
  <lpage>214</lpage>
</bibl>

<bibl id="B14">
  <title><p>{Identification of consensus patterns in unaligned DNA sequences
  known to be functionally related}</p></title>
  <aug>
    <au><snm>Hertz</snm><fnm>GZ</fnm></au>
    <au><snm>Hartzell</snm><fnm>III</fnm></au>
    <au><snm>Stormo</snm><fnm>GD</fnm></au>
  </aug>
  <source>Computational and Applied Biosciences</source>
  <pubdate>1990</pubdate>
  <volume>6</volume>
  <issue>2</issue>
  <fpage>81</fpage>
  <lpage>92</lpage>
</bibl>

<bibl id="B15">
  <title><p>Identifying DNA and protein patterns with statistically significant
  alignments of multiple sequences.</p></title>
  <aug>
    <au><snm>Hertz</snm><fnm>G Z</fnm></au>
    <au><snm>Stormo</snm><fnm>G D</fnm></au>
  </aug>
  <source>Bioinformatics</source>
  <pubdate>1999</pubdate>
  <volume>15</volume>
  <issue>7</issue>
  <fpage>563</fpage>
  <lpage>577</lpage>
</bibl>

<bibl id="B16">
  <title><p>DNA binding sites: representation and discovery</p></title>
  <aug>
    <au><snm>Stormo</snm><fnm>GD</fnm></au>
  </aug>
  <source>Bioinformatics</source>
  <pubdate>2000</pubdate>
  <volume>16</volume>
  <issue>1</issue>
  <fpage>16</fpage>
  <lpage>23</lpage>
</bibl>

<bibl id="B17">
  <title><p>Machine learning methods for transcription data
  integration</p></title>
  <aug>
    <au><snm>Holloway</snm><fnm>DT</fnm></au>
    <au><snm>Kon</snm><fnm>MA</fnm></au>
    <au><snm>DeLisi</snm><fnm>C</fnm></au>
  </aug>
  <source>IBM Journal of Research and Development</source>
  <pubdate>2006</pubdate>
  <volume>50</volume>
  <issue>6</issue>
  <fpage>631</fpage>
  <lpage>644</lpage>
</bibl>

<bibl id="B18">
  <title><p>SVMotif: A Machine Learning Motif Algorithm</p></title>
  <aug>
    <au><snm>Kon</snm><fnm>MA</fnm></au>
    <au><snm>Fan</snm><fnm>Y</fnm></au>
    <au><snm>Holloway</snm><fnm>D</fnm></au>
    <au><snm>DeLisi</snm><fnm>C</fnm></au>
  </aug>
  <source>ICMLA '07: Proceedings of the Sixth International Conference on
  Machine Learning and Applications</source>
  <publisher>Washington, DC, USA: IEEE Computer Society</publisher>
  <pubdate>2007</pubdate>
  <fpage>573</fpage>
  <lpage>-580</lpage>
</bibl>

<bibl id="B19">
  <title><p>{Large-Scale Mapping and Validation of Escherichia coli
  Transcriptional Regulation from a Compendium of Expression
  Profiles}</p></title>
  <aug>
    <au><snm>Faith</snm><fnm>JJ</fnm></au>
    <au><snm>Hayete</snm><fnm>B</fnm></au>
    <au><snm>Thaden</snm><fnm>JT</fnm></au>
    <au><snm>Mogno</snm><fnm>I</fnm></au>
    <au><snm>Wierzbowski</snm><fnm>J</fnm></au>
    <au><snm>Cottarel</snm><fnm>G</fnm></au>
    <au><snm>Kasif</snm><fnm>S</fnm></au>
    <au><snm>Collins</snm><fnm>JJ</fnm></au>
    <au><snm>Gardner</snm><fnm>TS</fnm></au>
  </aug>
  <source>PLoS Biology</source>
  <pubdate>2007</pubdate>
  <volume>5</volume>
  <issue>1</issue>
  <fpage>e8+</fpage>
</bibl>

<bibl id="B20">
  <title><p>{SIRENE: supervised inference of regulatory networks}</p></title>
  <aug>
    <au><snm>Mordelet</snm><fnm>F</fnm></au>
    <au><snm>Vert</snm><fnm>JP</fnm></au>
  </aug>
  <source>Bioinformatics</source>
  <pubdate>2008</pubdate>
  <volume>24</volume>
  <issue>16</issue>
  <fpage>76</fpage>
  <lpage>82</lpage>
</bibl>

<bibl id="B21">
  <title><p>Assessing computational tools for the discovery of transcription
  factor binding sites</p></title>
  <aug>
    <au><snm>Tompa</snm><fnm>M</fnm></au>
    <au><snm>Li</snm><fnm>N</fnm></au>
    <au><snm>Bailey</snm><fnm>TL</fnm></au>
    <au><snm>Church</snm><fnm>GM</fnm></au>
    <au><snm>De Moor</snm><fnm>B</fnm></au>
    <au><snm>Eskin</snm><fnm>E</fnm></au>
    <au><snm>Favorov</snm><fnm>AV</fnm></au>
    <au><snm>Frith</snm><fnm>MC</fnm></au>
    <au><snm>Fu</snm><fnm>Y</fnm></au>
    <au><snm>Kent</snm><fnm>WJ</fnm></au>
    <au><snm>Makeev</snm><fnm>VJ</fnm></au>
    <au><snm>Mironov</snm><fnm>AA</fnm></au>
    <au><snm>Noble</snm><fnm>WS</fnm></au>
    <au><snm>Pavesi</snm><fnm>G</fnm></au>
    <au><snm>Pesole</snm><fnm>G</fnm></au>
    <au><snm>Regnier</snm><fnm>M</fnm></au>
    <au><snm>Simonis</snm><fnm>N</fnm></au>
    <au><snm>Sinha</snm><fnm>S</fnm></au>
    <au><snm>Thijs</snm><fnm>G</fnm></au>
    <au><snm>Helden</snm><fnm>J</fnm></au>
    <au><snm>Vandenbogaert</snm><fnm>M</fnm></au>
    <au><snm>Weng</snm><fnm>Z</fnm></au>
    <au><snm>Workman</snm><fnm>C</fnm></au>
    <au><snm>Ye</snm><fnm>C</fnm></au>
    <au><snm>Zhu</snm><fnm>Z</fnm></au>
  </aug>
  <source>Nature Biotechnology</source>
  <pubdate>2005</pubdate>
  <volume>23</volume>
  <fpage>137</fpage>
  <lpage>144</lpage>
</bibl>

<bibl id="B22">
  <title><p>WebMOTIFS: automated discovery, filtering and scoring of DNA
  sequence motifs using multiple programs and Bayesian approaches</p></title>
  <aug>
    <au><snm>Romer</snm><fnm>KA</fnm></au>
    <au><snm>Kayombya</snm><fnm>GR</fnm></au>
    <au><snm>Fraenkel</snm><fnm>E</fnm></au>
  </aug>
  <source>Nucleic Acids Research</source>
  <pubdate>2007</pubdate>
  <volume>35</volume>
  <issue>suppl\_2</issue>
  <fpage>W217</fpage>
  <lpage>220</lpage>
</bibl>

<bibl id="B23">
  <title><p>{TAMO: a flexible, object-oriented framework for analyzing
  transcriptional regulation using DNA-sequence motifs}</p></title>
  <aug>
    <au><snm>Gordon</snm><fnm>DB</fnm></au>
    <au><snm>Nekludova</snm><fnm>L</fnm></au>
    <au><snm>McCallum</snm><fnm>S</fnm></au>
    <au><snm>Fraenkel</snm><fnm>E</fnm></au>
  </aug>
  <source>Bioinformatics</source>
  <pubdate>2005</pubdate>
  <volume>21</volume>
  <issue>14</issue>
  <fpage>3164</fpage>
  <lpage>3165</lpage>
</bibl>

<bibl id="B24">
  <title><p>Transcriptional regulatory code of a eukaryotic genome</p></title>
  <aug>
    <au><snm>Harbison</snm><fnm>CT.</fnm></au>
    <au><snm>Gordon</snm><fnm>DB</fnm></au>
    <au><snm>Lee</snm><fnm>TI</fnm></au>
    <au><snm>Rinaldi</snm><fnm>NJ</fnm></au>
    <au><snm>Macisaac</snm><fnm>KD</fnm></au>
    <au><snm>Danford</snm><fnm>TW</fnm></au>
    <au><snm>Hannett</snm><fnm>NM</fnm></au>
    <au><snm>Tagne</snm><fnm>JB</fnm></au>
    <au><snm>Reynolds</snm><fnm>DB</fnm></au>
    <au><snm>Yoo</snm><fnm>J</fnm></au>
    <au><snm>Jennings</snm><fnm>EG</fnm></au>
    <au><snm>Zeitlinger</snm><fnm>J</fnm></au>
    <au><snm>Pokholok</snm><fnm>DK</fnm></au>
    <au><snm>Kellis</snm><fnm>M</fnm></au>
    <au><snm>Rolfe</snm><fnm>PA</fnm></au>
    <au><snm>Takusagawa</snm><fnm>KT</fnm></au>
    <au><snm>Lander</snm><fnm>ES</fnm></au>
    <au><snm>Gifford</snm><fnm>DK</fnm></au>
    <au><snm>Fraenkel</snm><fnm>E</fnm></au>
    <au><snm>Young</snm><fnm>RA</fnm></au>
  </aug>
  <source>Nature</source>
  <pubdate>2004</pubdate>
  <volume>431</volume>
  <issue>7004</issue>
  <fpage>99</fpage>
  <lpage>104</lpage>
</bibl>

<bibl id="B25">
  <title><p>EMD: an ensemble algorithm for discovering regulatory motifs in DNA
  sequences</p></title>
  <aug>
    <au><snm>Hu</snm><fnm>J</fnm></au>
    <au><snm>Yang</snm><fnm>Y</fnm></au>
    <au><snm>Kihara</snm><fnm>D</fnm></au>
  </aug>
  <source>BMC Bioinformatics</source>
  <pubdate>2006</pubdate>
  <volume>7</volume>
  <issue>1</issue>
  <fpage>342</fpage>
</bibl>

<bibl id="B26">
  <title><p>{Positional clustering improves computational binding site
  detection and identifies novel cis-regulatory sites in mammalian GABAA
  receptor subunit genes}</p></title>
  <aug>
    <au><cnm>Reddy</cnm></au>
    <au><snm>Timothy</snm><fnm>E.</fnm></au>
    <au><cnm>Shakhnovich</cnm></au>
    <au><snm>Boris</snm><fnm>E.</fnm></au>
    <au><cnm>Roberts</cnm></au>
    <au><snm>Daniel</snm><fnm>S.</fnm></au>
    <au><cnm>Russek</cnm></au>
    <au><snm>Shelley</snm><fnm>J.</fnm></au>
    <au><cnm>Delisi</cnm></au>
    <au><cnm>Charles</cnm></au>
  </aug>
  <source>Nucleic Acids Research</source>
  <publisher>Oxford University Press</publisher>
  <pubdate>2007</pubdate>
  <volume>35</volume>
  <issue>3</issue>
  <fpage>e20</fpage>
</bibl>

<bibl id="B27">
  <title><p>{Binding Site Graphs: A New Graph Theoretical Framework for
  Prediction of Transcription Factor Binding Sites}</p></title>
  <aug>
    <au><snm>Reddy</snm><fnm>TE</fnm></au>
    <au><snm>Delisi</snm><fnm>C</fnm></au>
    <au><snm>Shakhnovich</snm><fnm>BE</fnm></au>
  </aug>
  <source>PLoS Computational Biology</source>
  <pubdate>2007</pubdate>
  <volume>3</volume>
  <issue>5</issue>
  <fpage>e90+</fpage>
</bibl>

<bibl id="B28">
  <title><p>{M are better than one: an ensemble-based motif finder and its
  application to regulatory element prediction}</p></title>
  <aug>
    <au><snm>Yanover</snm><fnm>C</fnm></au>
    <au><snm>Singh</snm><fnm>M</fnm></au>
    <au><snm>Zaslavsky</snm><fnm>E</fnm></au>
  </aug>
  <source>Bioinformatics</source>
  <pubdate>2009</pubdate>
  <volume>25</volume>
  <issue>7</issue>
  <fpage>868</fpage>
  <lpage>874</lpage>
</bibl>

<bibl id="B29">
  <title><p>{In silico representation and discovery of transcription factor
  binding sites}</p></title>
  <aug>
    <au><snm>Pavesi</snm><fnm>G</fnm></au>
    <au><snm>Mauri</snm><fnm>G</fnm></au>
    <au><snm>Pesole</snm><fnm>G</fnm></au>
  </aug>
  <source>Brief Bioinformatics</source>
  <pubdate>2004</pubdate>
  <volume>5</volume>
  <fpage>217</fpage>
  <lpage>236</lpage>
</bibl>

<bibl id="B30">
  <title><p>BEST: Binding-site Estimation Suite of Tools</p></title>
  <aug>
    <au><snm>Che</snm><fnm>D</fnm></au>
    <au><snm>Jensen</snm><fnm>ST</fnm></au>
    <au><snm>Cai</snm><fnm>L</fnm></au>
    <au><snm>Liu</snm><fnm>JS</fnm></au>
  </aug>
  <source>Bioinformatics</source>
  <pubdate>2005</pubdate>
  <volume>21</volume>
  <issue>12</issue>
  <fpage>2909</fpage>
  <lpage>2911</lpage>
</bibl>

<bibl id="B31">
  <title><p>{GAME: detecting cis-regulatory elements using a genetic
  algorithm}</p></title>
  <aug>
    <au><snm>Wei</snm><fnm>Z</fnm></au>
    <au><snm>Jensen</snm><fnm>ST</fnm></au>
  </aug>
  <source>Bioinformatics</source>
  <pubdate>2006</pubdate>
  <volume>22</volume>
  <issue>13</issue>
  <fpage>1577</fpage>
  <lpage>1584</lpage>
</bibl>

<bibl id="B32">
  <title><p>{BioOptimizer: a Bayesian scoring function approach to motif
  discovery}</p></title>
  <aug>
    <au><snm>Jensen</snm><fnm>ST</fnm></au>
    <au><snm>Liu</snm><fnm>JS</fnm></au>
  </aug>
  <source>Bioinformatics</source>
  <pubdate>2004</pubdate>
  <volume>20</volume>
  <issue>10</issue>
  <fpage>1557</fpage>
  <lpage>1564</lpage>
</bibl>

<bibl id="B33">
  <title><p>{Locating mammalian transcription factor binding sites: A survey of
  computational and experimental techniques}</p></title>
  <aug>
    <au><snm>Elnitski</snm><fnm>L</fnm></au>
    <au><snm>Jin</snm><fnm>VX</fnm></au>
    <au><snm>Farnham</snm><fnm>PJ</fnm></au>
    <au><snm>Jones</snm><fnm>SJM</fnm></au>
  </aug>
  <source>Genome Research</source>
  <publisher>Genomic Functional Analysis Section, National Human Genome
  Research Institute, National Institutes of Health, Rockville, Maryland 20878,
  USA;</publisher>
  <pubdate>2006</pubdate>
  <volume>16</volume>
  <issue>12</issue>
  <fpage>1455</fpage>
  <lpage>-1464</lpage>
</bibl>

<bibl id="B34">
  <title><p>{An algorithm for finding signals of unknown length in DNA
  sequences}</p></title>
  <aug>
    <au><snm>Pavesi</snm><fnm>G</fnm></au>
    <au><snm>Mauri</snm><fnm>G</fnm></au>
    <au><snm>Pesole</snm><fnm>G</fnm></au>
  </aug>
  <source>Bioinformatics</source>
  <publisher>Department of Computer Science, Systems and Communication,
  University of Milan-Bicocca, Via Bicocca degli Arcimboldi 8, Milan, I-20126,
  Italy. pavesi@disco.unimib.it</publisher>
  <pubdate>2001</pubdate>
  <volume>17</volume>
  <issue>suppl\_1</issue>
  <fpage>S207</fpage>
  <lpage>-214</lpage>
</bibl>

<bibl id="B35">
  <title><p>{In silico regulatory analysis for exploring human disease
  progression}</p></title>
  <aug>
    <au><snm>Holloway</snm><fnm>D</fnm></au>
    <au><snm>Kon</snm><fnm>M</fnm></au>
    <au><snm>DeLisi</snm><fnm>C</fnm></au>
  </aug>
  <source>Biology Direct</source>
  <pubdate>2008</pubdate>
  <volume>3</volume>
  <issue>1</issue>
  <fpage>24+</fpage>
</bibl>

<bibl id="B36">
  <title><p>Discriminative motif discovery in DNA and protein sequences using
  the DEME algorithm</p></title>
  <aug>
    <au><snm>Redhead</snm><fnm>E</fnm></au>
    <au><snm>Bailey</snm><fnm>T</fnm></au>
  </aug>
  <source>BMC Bioinformatics</source>
  <pubdate>2007</pubdate>
  <volume>8</volume>
  <issue>1</issue>
  <fpage>385</fpage>
</bibl>

<bibl id="B37">
  <title><p>{Identification and analysis of functional elements in 1\% of the
  human genome by the ENCODE pilot project}</p></title>
  <aug>
    <au><snm>Consortium</snm><fnm>TEP</fnm></au>
  </aug>
  <source>Nature</source>
  <publisher>Nature Publishing Group</publisher>
  <pubdate>2007</pubdate>
  <volume>447</volume>
  <issue>7146</issue>
  <fpage>799</fpage>
  <lpage>-816</lpage>
</bibl>

<bibl id="B38">
  <title><p>{Rank order metrics for quantifying the association of sequence
  features with gene regulation}</p></title>
  <aug>
    <au><snm>Clarke</snm><fnm>ND</fnm></au>
    <au><snm>Granek</snm><fnm>JA</fnm></au>
  </aug>
  <source>Bioinformatics</source>
  <pubdate>2003</pubdate>
  <volume>19</volume>
  <issue>2</issue>
  <fpage>212</fpage>
  <lpage>218</lpage>
</bibl>

<bibl id="B39">
  <title><p>Ensemble Machine Methods for DNA Binding</p></title>
  <aug>
    <au><snm>Fan</snm><fnm>Y</fnm></au>
    <au><snm>Kon</snm><fnm>MA</fnm></au>
    <au><snm>DeLisi</snm><fnm>C</fnm></au>
  </aug>
  <source>Proceedings of the 2008 Seventh International Conference on Machine
  Learning and Applications</source>
  <publisher>Washington, DC, USA: IEEE Computer Society</publisher>
  <pubdate>2008</pubdate>
  <fpage>709</fpage>
  <lpage>-716</lpage>
</bibl>

<bibl id="B40">
  <title><p>{Transcriptional Regulatory Networks in Saccharomyces
  cerevisiae}</p></title>
  <aug>
    <au><snm>Lee</snm><fnm>TI</fnm></au>
    <au><snm>Rinaldi</snm><fnm>NJ</fnm></au>
    <au><snm>Robert</snm><fnm>F</fnm></au>
    <au><snm>Odom</snm><fnm>DT</fnm></au>
    <au><snm>Bar Joseph</snm><fnm>Z</fnm></au>
    <au><snm>Gerber</snm><fnm>GK</fnm></au>
    <au><snm>Hannett</snm><fnm>NM</fnm></au>
    <au><snm>Harbison</snm><fnm>CT</fnm></au>
    <au><snm>Thompson</snm><fnm>CM</fnm></au>
    <au><snm>Simon</snm><fnm>I</fnm></au>
    <au><snm>Zeitlinger</snm><fnm>J</fnm></au>
    <au><snm>Jennings</snm><fnm>EG</fnm></au>
    <au><snm>Murray</snm><fnm>HL</fnm></au>
    <au><snm>Gordon</snm><fnm>DB</fnm></au>
    <au><snm>Ren</snm><fnm>B</fnm></au>
    <au><snm>Wyrick</snm><fnm>JJ</fnm></au>
    <au><snm>Tagne</snm><fnm>JB</fnm></au>
    <au><snm>Volkert</snm><fnm>TL</fnm></au>
    <au><snm>Fraenkel</snm><fnm>E</fnm></au>
    <au><snm>Gifford</snm><fnm>DK</fnm></au>
    <au><snm>Young</snm><fnm>RA</fnm></au>
  </aug>
  <source>Science</source>
  <pubdate>2002</pubdate>
  <volume>298</volume>
  <issue>5594</issue>
  <fpage>799</fpage>
  <lpage>804</lpage>
</bibl>

<bibl id="B41">
  <title><p>{The UCSC Genome Browser database: update 2010}</p></title>
  <aug>
    <au><snm>Rhead</snm><fnm>B</fnm></au>
    <au><snm>Karolchik</snm><fnm>D</fnm></au>
    <au><snm>Kuhn</snm><fnm>RM</fnm></au>
    <au><snm>Hinrichs</snm><fnm>AS</fnm></au>
    <au><snm>Zweig</snm><fnm>AS</fnm></au>
    <au><snm>Fujita</snm><fnm>PA</fnm></au>
    <au><snm>Diekhans</snm><fnm>M</fnm></au>
    <au><snm>Smith</snm><fnm>KE</fnm></au>
    <au><snm>Rosenbloom</snm><fnm>KR</fnm></au>
    <au><snm>Raney</snm><fnm>BJ</fnm></au>
    <au><snm>Pohl</snm><fnm>A</fnm></au>
    <au><snm>Pheasant</snm><fnm>M</fnm></au>
    <au><snm>Meyer</snm><fnm>LR</fnm></au>
    <au><snm>Learned</snm><fnm>K</fnm></au>
    <au><snm>Hsu</snm><fnm>F</fnm></au>
    <au><snm>Hillman Jackson</snm><fnm>J</fnm></au>
    <au><snm>Harte</snm><fnm>RA</fnm></au>
    <au><snm>Giardine</snm><fnm>B</fnm></au>
    <au><snm>Dreszer</snm><fnm>TR</fnm></au>
    <au><snm>Clawson</snm><fnm>H</fnm></au>
    <au><snm>Barber</snm><fnm>GP</fnm></au>
    <au><snm>Haussler</snm><fnm>D</fnm></au>
    <au><snm>Kent</snm><fnm>WJ</fnm></au>
  </aug>
  <source>Nucleic Acids Research</source>
  <pubdate>2010</pubdate>
  <volume>38</volume>
  <issue>suppl\_1</issue>
  <fpage>D613</fpage>
  <lpage>619</lpage>
</bibl>

<bibl id="B42">
  <title><p>Probabilistic outputs for support vector machines and comparison to
  regularized likelihood methods</p></title>
  <aug>
    <au><snm>Platt</snm><fnm>J.</fnm></au>
  </aug>
  <editor>A.J. Smola, P. Bartlett, B. Schoelkopf, D. Schuurmans</editor>
  <pubdate>2000</pubdate>
  <fpage>61</fpage>
  <lpage>-74</lpage>
</bibl>

<bibl id="B43">
  <title><p>TRANSFAC: a database on transcription factors and their DNA binding
  sites.</p></title>
  <aug>
    <au><snm>Wingender</snm><fnm>E.</fnm></au>
    <au><snm>Dietze</snm><fnm>P.</fnm></au>
    <au><snm>Karas</snm><fnm>H.</fnm></au>
    <au><snm>Kn\"{u}ppel</snm><fnm>R.</fnm></au>
  </aug>
  <source>Nucleic Acids Research</source>
  <pubdate>1996</pubdate>
  <volume>24</volume>
  <issue>1</issue>
  <fpage>238</fpage>
  <lpage>-241</lpage>
</bibl>

<bibl id="B44">
  <title><p>{An improved map of conserved regulatory sites for Saccharomyces
  cerevisiae}</p></title>
  <aug>
    <au><snm>MacIsaac</snm><fnm>K</fnm></au>
    <au><snm>Wang</snm><fnm>T</fnm></au>
    <au><snm>Gordon</snm><fnm>DB</fnm></au>
    <au><snm>Gifford</snm><fnm>D</fnm></au>
    <au><snm>Stormo</snm><fnm>G</fnm></au>
    <au><snm>Fraenkel</snm><fnm>E</fnm></au>
  </aug>
  <source>BMC Bioinformatics</source>
  <pubdate>2006</pubdate>
  <volume>7</volume>
  <issue>1</issue>
  <fpage>113+</fpage>
</bibl>

<bibl id="B45">
  <title><p>{JASPAR: an open-access database for eukaryotic transcription
  factor binding profiles.}</p></title>
  <aug>
    <au><snm>Sandelin</snm><fnm>A</fnm></au>
    <au><snm>Alkema</snm><fnm>W</fnm></au>
    <au><snm>Engstr\"{o}m</snm><fnm>P</fnm></au>
    <au><snm>Wasserman</snm><fnm>WW</fnm></au>
    <au><snm>Lenhard</snm><fnm>B</fnm></au>
  </aug>
  <source>Nucleic acids research</source>
  <pubdate>2004</pubdate>
  <volume>32</volume>
  <issue>Database issue</issue>
  <fpage>91</fpage>
  <lpage>94</lpage>
</bibl>

<bibl id="B46">
  <title><p>Gene selection for cancer classification using support vector
  machines</p></title>
  <aug>
    <au><snm>Guyon</snm><fnm>I</fnm></au>
    <au><snm>Weston</snm><fnm>J</fnm></au>
    <au><snm>Barnhill</snm><fnm>S</fnm></au>
    <au><snm>Vapnik</snm><fnm>V</fnm></au>
  </aug>
  <source>Machine Learning</source>
  <pubdate>2002</pubdate>
  <volume>46</volume>
  <issue>1</issue>
  <fpage>389</fpage>
  <lpage>422</lpage>
</bibl>

<bibl id="B47">
  <title><p>Recursive SVM feature selection and sample classification for
  mass-spectrometry and microarray data</p></title>
  <aug>
    <au><snm>Zhang</snm><fnm>X</fnm></au>
    <au><snm>Lu</snm><fnm>X</fnm></au>
    <au><snm>Shi</snm><fnm>Q</fnm></au>
    <au><snm>Xu</snm><fnm>X</fnm></au>
    <au><snm>E. Leung</snm><fnm>H</fnm></au>
    <au><snm>Harris</snm><fnm>LN</fnm></au>
    <au><snm>Iglehart</snm><fnm>JD</fnm></au>
    <au><snm>Miron</snm><fnm>A</fnm></au>
    <au><snm>Liu</snm><fnm>JS</fnm></au>
    <au><snm>Wong</snm><fnm>WH</fnm></au>
  </aug>
  <source>BMC Bioinformatics</source>
  <pubdate>2006</pubdate>
  <volume>7</volume>
  <issue>1</issue>
  <fpage>197</fpage>
</bibl>

</refgrp>
} 

\newpage
{\ifthenelse{\boolean{publ}}{\footnotesize}{\small}
 \bibliographystyle{bmc_article}  
  \bibliography{AllYF4} }     


\ifthenelse{\boolean{publ}}{\end{multicols}}{}

\end{bmcformat}
\end{document}